\def\spose#1{\hbox to 0pt{#1\hss}}
\newcommand\simlt{\mathrel{\spose{\lower 3pt\hbox{$\mathchar"218$}}
     \raise 2.0pt\hbox{$\mathchar"13C$}}}
\newcommand\simgt{\mathrel{\spose{\lower 3pt\hbox{$\mathchar"218$}}
     \raise 2.0pt\hbox{$\mathchar"13E$}}}

\documentclass[useAMS,usenatbib]{mn2e}
\usepackage{psfig}
\usepackage{epsfig}
\usepackage{lscape}
\usepackage{amssymb}
\defcitealias{Bro07}{Paper\,I}
\defcitealias{Bry08}{Paper\,III}
\defcitealias{McC96}{McCarthy et al. (1996; 2007, private communication)}

\title[ Distant radio galaxies in the Southern hemisphere 
-- II. ]{
A new search for distant radio galaxies in the Southern hemisphere -- II. 2.2$\mu$m imaging}
\author[J. J. Bryant et al.]{J. J. Bryant$^{1}$\thanks{E-mail:
jbryant@physics.usyd.edu.au},  J. W. Broderick$^{1}$, H. M. Johnston$^{1}$, 
R. W. Hunstead$^{1}$, \newauthor B. M. Gaensler$^{1,2}$, and C. De Breuck$^{3}$ 
\\
$^{1}$Institute of Astronomy, School of Physics A29, The University of Sydney, NSW 2006, Australia\\
$^{2}$Harvard-Smithsonian Center for Astrophysics, 60 Garden Street, Cambridge MA 02138, USA\\
$^{3}$European Southern Observatory, Karl Schwarzschild Stra\ss e 2, D-85748 Garching, Germany}
\begin{document}

\date{}

\pagerange{\pageref{firstpage}--\pageref{lastpage}} \pubyear{2006}

\maketitle

\label{firstpage}

\begin{abstract}

We have compiled a sample of 234 ultra-steep-spectrum (USS) selected radio
sources in order to find high-redshift radio
galaxies.
The sample covers the declination range $-40^{\circ} < \delta < -30^{\circ}$
in
the overlap region between the 1400-MHz NRAO VLA Sky Survey, 408-MHz Revised Molonglo Reference
Catalogue and the 843-MHz Sydney University Molonglo Sky Survey (the
MRCR--SUMSS sample). This is the second in a series of papers on the
MRCR--SUMSS sample, and here we present the $K$-band (2.2$\mu$m) imaging of 173 of the
sources primarily from 
the Magellan 
and
the Anglo-Australian Telescopes. 
We detect a counterpart to the radio source
in 93 per cent of the new
$K$-band images which, along with previously published data, makes this the largest published sample of $K$-band
counterparts to USS-selected radio galaxies.
The location of the $K$-band identification has been compared to the
features of the radio emission for the double sources.
We find that the
identification is most likely to lie near the midpoint of the radio lobes
rather than closer to the brighter lobe, making the centroid a less likely
place to find the optical counterpart.
79 per cent of the identifications are less than 1\,arcsec
from the radio lobe axis. These results differ from studies of low-redshift radio 
samples where the environments are typically not nearly
so dense and disturbed as those at high redshift.
In contrast to some literature samples, we find that the majority
of our sample shows no alignment between the near-infrared and radio axes.
Several different morphologies of aligned structures are found
and those that are aligned within 10\,degrees are consistent with jet-induced star formation.
The distribution and median value of the $K$-band magnitudes for the MRCR--SUMSS sample are found to
be similar to several other USS-selected samples even though each sample
has a very different median 1400\,MHz flux density. USS-selection from a lower-radio-frequency sample
has not netted fainter $K$-band magnitudes, which may imply that the k-correction is not
responsible for the effectiveness of USS-selection.
Due to space constraints, lower resolution figures are shown here, and the full 
resolution is available in the MNRAS published paper.

\end{abstract}

\begin{keywords}
galaxies: active -- surveys -- galaxies: high-redshift -- infrared: galaxies -- radio continuum: galaxies.
\end{keywords}

\section{INTRODUCTION}
\label{Intro}

At low-redshift, powerful radio galaxies are hosted by
giant ellipticals \citep{mat64}.
The inference
that high-redshift radio galaxies (HzRGs; $z>2$) host a supermassive ($>10^{9}$ M$_{\odot}$) black hole
\citep{Bla82}, and the close connection between the mass
of the black hole and that of the galaxy \citep[e.g.][]{geb03},
provides an evolutionary path to the present-day giant ellipticals.
HzRGs are therefore excellent laboratories for studying the
evolution of massive galaxies.

Most HzRGs found to date have been
selected on the basis of their steep radio spectral index $\alpha$ (where
$S_\nu \propto \nu^{\alpha}$ for flux density $S$ at frequency $\nu$), since ultra-steep-spectrum
(USS; $\alpha < -1.0$) radio galaxies are more likely to be found
at $z>2$
\citep{BM:79,Tie:79,deB00,deB06}. The physical basis for this
so-called $z-\alpha$ correlation was previously thought to be due
to a concave radio galaxy spectrum that was assumed to steepen to higher
frequencies. The steeper part of the spectrum would be redshifted, giving a steeper
spectral index at a given observed frequency for a high-redshift source than a low-redshift one.
However, this interpretation has recently been challenged
by \citet{kla06} who introduced a new paradigm
for interpreting the radio spectra of high-redshift massive galaxies,
in which USS massive galaxies are located
in extremely dense environments, comparable to rich clusters of
galaxies in the local Universe. The high ambient gas density in this
environment means that the radio lobes are pressure-confined and lose
their energy slowly by synchrotron and inverse Compton losses (rather
than rapidly by adiabatic expansion as occurs in more rarefied
environments at low redshift). If this interpretation is correct, then
not only do radio galaxies represent the extreme limit of the galaxy
mass function at early times, they may also locate
the most extreme overdensities of matter in the early Universe.

The number density of HzRGs is low, only a few times
$10^{-8}$\,Mpc$^{-3}$ 
\citep{Ven:07}.
Of these, 96 per cent are detected to $K=22$ \citep{Mil:08} and hence
a $K$-band survey is an efficient method of identifying the
hosts of USS radio-selected galaxies.
The spectral energy distribution of old ($>1$\,Gyr) stars
peaks in the near-infrared at low redshift, while that of young blue
stars dominates the $K$-band at high redshift, and therefore it has been 
argued that the $K$-band 
emission can be assumed to trace the stellar mass.
HzRGs have the highest $K$-band luminosities in the early Universe, 
and are brighter than
optically-selected galaxies at all redshifts,
which demonstrates that radio galaxies are among the most massive
galaxies at all epochs \citep{roc04,Sey07}. 
Furthermore, they follow a well-known linear\footnote{Linear fit
between $\log(z)$ and $K$ magnitude corresponds to a power-law
relation between $z$ and $2.2\mu$m flux.} correlation between 
redshift and $K$ magnitude, the $K$--$z$ relation \citep{Lil:84}, 
which is an efficient tool for filtering out
the low-redshift galaxies, leaving a smaller sample of potential high-redshift objects for follow-up spectroscopy.

We have carried out a search in the Southern Hemisphere
designed to find high-redshift ($z>2$) radio galaxies, to
study their evolution, and to
form a significant sample of those at $2<z<3.5$ for a
detailed quantitative study of their environments.
This is the first survey in the South to use sensitive low-frequency catalogues
to data mine a large area of sky, and it became possible with the completion of
the 843-MHz Sydney University Molonglo Sky Survey \citep[SUMSS;][]{Boc99,Mau03} and the reanalysis of
the 408-MHz Molonglo cross data \citep{Lar81} 
to give the Revised Molonglo Reference Catalogue (MRCR; Crawford, private communication).

The details of the sample selection and the radio imaging results were
given in \citet[][hereafter Paper\,I]{Bro07}.
The present paper discusses the $K$-band identifications and magnitudes for
173 of the radio sources in \citetalias{Bro07}, along with new
higher-resolution radio images. The optical spectra for
sources from the MRCR--SUMSS sample are introduced in \citet[][hereafter Paper\,III]{Bry08} 
which 
presents a detailed analysis of the radio, $K$-band and redshift
information for the sample. 
We have adopted a flat, $\Lambda$ cold dark matter cosmology with
$H_0=71$ km s$^{-1}$ Mpc$^{-1}$, $\Omega_{\rm M}=0.27$ and $\Omega_{\Lambda}=0.73$.

\section{Observations and data reduction}
\subsection{Target selection}

The sources for our survey were selected by cross-matching 

\noindent (i) the 843-MHz SUMSS catalogue,

\noindent (ii) the 408\,MHz MRCR, which lowered the
earlier MRC flux density limit of 0.95\,Jy to  $\sim$200\,mJy
($5\sigma$), and

\noindent (iii) the 1400-MHz NVSS \citep{Con98} in the overlap region 
with SUMSS (declination $-30^{\circ}$ to $-40^{\circ}$).

We then selected sources with a spectral index between 408 to 843\,MHz, $\alpha_{408}^{843}<-1.0$ and
$|b|>10^{\circ}$
(to avoid confusion from the Galactic plane). 
Double lobed sources were inspected by eye to ensure that
both components were included in the flux density measure.
Minimum flux density cut-offs of 200mJy and 15mJy were applied to the
MRCR and NVSS catalogues respectively to improve the accuracy of the
spectral indices. 

The target selection is discussed in detail in \citetalias{Bro07}.

\subsection{Radio observations - ATCA}

\subsubsection{1384- and 2368-MHz imaging}
Radio images were obtained with the Australia Telescope Compact Array (ATCA)
in several runs in 2003--2006. Both 1384- and 2368-MHz
images were needed to
pinpoint the $K$-band identification. 
The full details
of the radio observations and reduction are given in \citetalias{Bro07} where the 
analysis was done with natural-weighted images. However, in this paper, in 
cases where the $K$-band counterpart was 
not clear, we used
uniform-weighted images as they have better
resolution of $\sim6.1 \times 3.5$\,arcsec$^{2}$ at 2368\,MHz.

\subsubsection{4800- and 8640-MHz imaging}
Higher-resolution observations were obtained for 
29 sources on {\footnotesize UT} 2008 January 16 -- 20.
Dual-frequency observations at
4800 and 8640 MHz were carried out with a $2\times 128$ MHz bandwidth correlator
configuration. The 6A array configuration was used, spanning baselines from 337 to
5939 m.  The median integration time per source was $\sim$70 min,
consisting of well-spaced cuts of 3--4 min in duration. PKS B1934$-$638
was used as the primary calibrator, while phase calibrators
were observed every 20
min. A log of observations for the 4800- and 8640-MHz data is given in Table~\ref{obslog}.

The data were reduced with {\sc miriad} \citep{Sau95}, following a procedure similar to that described in \citetalias{Bro07}.
We used natural weighting to optimise the sensitivity; the median angular
resolution is $4.2 \times 2.9$ arcsec$^{2}$ at 4800 MHz and $2.3 \times
1.6$ arcsec$^{2}$ at 8640 MHz. {\footnotesize CLEAN} boxes were defined
using both the automatically constrained {\footnotesize CLEAN} technique
described in \citetalias{Bro07}, and manual selection where necessary. 
After the {\footnotesize CLEAN} regions had been defined, each dirty image was 
{\footnotesize CLEAN}ed to 2--3 times the theoretical rms noise level; typically 
only $\sim$100 iterations were necessary.
Phase-only self-calibration was then used to improve the
dynamic range. The median rms noise levels are 0.14 mJy\,beam$^{-1}$ at
4800 MHz and 0.20 mJy\,beam$^{-1}$ at 8640 MHz; the measured rms noise 
level is in close agreement with the theoretical value in all but a 
handful of the {\footnotesize CLEAN} images.
We estimate that the typical positional uncertainty is $<0.5$\,arcsec
at both 4800 and 8640\,MHz. However, in a handful of cases, the positional
uncertainty may be up to $\sim$0.5\,arcsec at 8640\,MHz because of poor
phase stability.

\subsection{$K$-band ($2.2\mu$m) observations}

The $K$-band observations are listed in the log of observations in Table~\ref{obslog}.
A $Ks$ filter was used for the $K$-band imaging.

\begin{table*}
\caption{Log of observations.}
\label{obslog}
\begin{center}
\begin{tabular}{lccc}
\hline
\hline
Date & Telescope & Instrument & Optical seeing\\
  &           &             & (arcsec)   \\
\hline
\multicolumn{3}{l}{{\bf $K$-band imaging:}} &  \\
 2004 August 3--5 & AAT & IRIS2 &  0.9--1.9 \\
 2005 June 21--24 & AAT & IRIS2  & 0.9--1.9 \\
 2006 June 11--13 & Magellan & PANIC  & 0.5--1.5 \\
 2006 November 8--10 &  Magellan & PANIC   & 0.5--1.3  \\
 2007 April 2--4 &  Magellan & PANIC  & 0.5--1.0  \\
\multicolumn{3}{l}{{\bf 4800- \& 8640-MHz imaging:}}  & \\
 2008 January 16--20 & ATCA &  &  \\
\hline
\end{tabular}
\end{center}
\end{table*}

\begin{figure}
\psfig{file=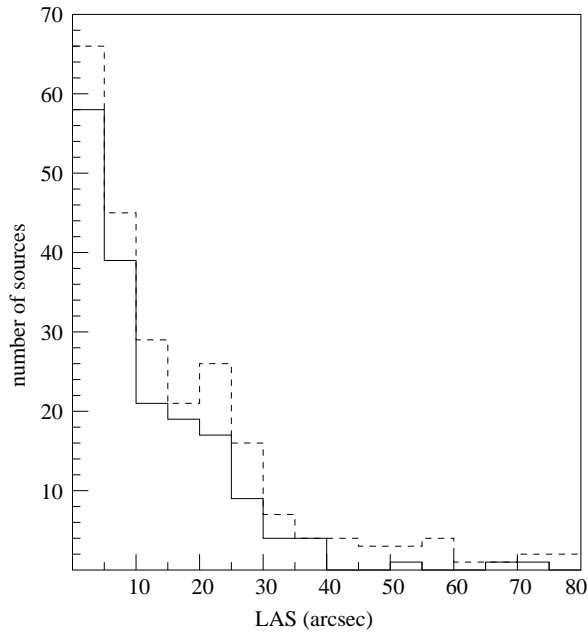, width=7.7cm}
\caption{Distribution of the largest angular size from the 2368-MHz images (or 4800- or 8640-MHz 
images where
available) for all of the MRCR--SUMSS sources from \citetalias{Bro07} (dashed line) and from Table~\ref{data} for the objects
that were imaged in $K$-band (solid line).
} 
\label{LAShist}
\end{figure}

Of the 234 sources with radio imaging, two already have 2MASS magnitudes and 
we have selected 
a further 163 sources for $K$-band imaging. In addition, there are eight 
targets in our sample that have previously been observed by \citet{deB04}.
The selection of sources for $K$-band followup was primarily based on the Right Ascensions
accessible during the observing sessions, and hence there is no bias
in the $K$-band sample compared to the radio sample. There was no selection
made on the basis of flux density or radio morphology.
Fig.~\ref{LAShist} shows that
the sources chosen for $K$-band imaging are representative of the largest angular size (LAS)
distribution of the whole radio sample.

\subsubsection{IRIS2}

Our initial $Ks$-band imaging was begun on the 3.9-m Anglo-Australian
Telescope's (AAT) IRIS2 detector \citep{gil00}.
With a
scale of 0.446\,arcsec\,pixel$^{-1}$, the $1024 \times 1024$\,pixel HAWAII HgCdTe array
imaged a sky area of $8 \times 8$ arcmin$^{2}$.
Two observing runs on 2004 August 3--5 and 2005 June 21--24
were mostly
clouded out. We did, however, obtain $Ks$-band images of 42 sources in seeing
of 0.9--1.9\,arcsec.
Integration times were 15\,mins, made up of 8 cycles of 8\,s
exposures in a 14-point random jitter pattern.
There were seven non-detections, of which five were reobserved on the
larger
Magellan telescope, as discussed in Section~\ref{PANICobs}.

The data were reduced with the pipeline data reduction package, {\sc oracdr}, which sky-subtracted, flat-fielded, shifted and combined
the individual dithered images.
The faintest objects measured had $Ks=19.6$.

\subsubsection{PANIC}
\label{PANICobs}

We observed 126 targets in $Ks$-band with Persson's Auxiliary Nasmyth Infrared Camera 
\citep[PANIC;][]{mar04} 
on the 6.5-m Magellan Baade telescope at Las Campanas Observatory
on 2006 June 11--13, 2006 November 8--10 and 2007 April 2--4.  These included
five objects that remained
undetected in IRIS2 images and two targets which were undetected in
previous NTT observations \citep[see][]{deB04}. The field of
view is $2\times2$\,arcmin$^{2}$ from the $1024\times 1024$ pixel HgCdTe Hawaii
detector with $0.125$\,arcsec\,pixel$^{-1}$. Seeing in the optical ranged from 
0.6--0.8\,arcsec for most of the three
observing runs,
only briefly exceeding 1.0\,arcsec, and at best reaching just under
$0.5$\,arcsec. The conditions were
photometric apart from two hours of thin cirrus in the 2006 June run.
We used a 9-point dither pattern of 20-s exposures and
repeated this pattern with three loops giving an exposure
time of 9\,mins. Reduction was done in real time and if the object was not
detected in 9\,mins, the integration was repeated.

We had intermittent problems with moon glow causing slight to very strong
stripes across the image. This was mostly due to reflection of the moon off the
telescope baffles. In the 2006 November run we had brief periods of
interference in the images due to an electronics problem. Both of these
issues were short lived and when they occurred they affected only some of the
nine dithered images. In those cases, the nine images were carefully inspected,
and only the clean dithers were combined. Hence, some of the integration times
for PANIC images are not multiples of 9\,mins (see Table~\ref{data}).

Reduction was done using the {\sc gopanic} reduction pipeline
routine, which is part of the {\sc panic} package in {\sc iraf} by P. Martini.
The pipeline summed the individual exposures, applied a linearity
correction, flat fielded and sky subtracted the images, did a distortion 
correction
then shifted and added the individual dithered images into a final frame. 
When multiple dither
sequences were done, the shifts measured in {\sc iraf centers} were used in
{\sc iraf imcombine} to make the final image. The deepest images reached
$Ks=21.5$ (in just under 1\,hr integration). Counterparts for nine sources
were not detected.

\subsubsection{Astrometry}

Accurate coordinates were applied to the final IRIS2 and PANIC images
by comparison with the SuperCOSMOS Sky Survey \citep{Ham01} 
images 
using the {\sc karma}\footnote{http://www.atnf.csiro.au/computing/software/karma/} 
astrometry package \citep{goo96} 
{\sc koords}. The SuperCOSMOS astrometry is accurate to
$\leq0.2$-arcsec rms. 
There are several fields at high Galactic latitude with 
larger astrometric uncertainties;
these fields have as few as three stars in the
PANIC field of view. The majority of fields had more than seven stars suitable
for astrometry.

The astrometric accuracy
on the $Ks$-band images was sufficient in most cases to identify the counterpart
to each radio source.
Of the 11 sources (9 from PANIC and 2 from IRIS2) without a $K$-band identification, it is most likely
that the identifications are simply fainter than the image limit or the radio structures are extended with
ambiguous identifications. The non-detections are discussed further in 
Section~\ref{notes}. We are confident that the
non-detections are not due to astrometric misalignment. 

\subsubsection{$K$-band photometry}

Two of the objects already had $K$-band magnitudes listed in the Two
Micron All Sky Survey \citep[2MASS;][]{Skr06} and therefore did not need to be reobserved.
Both the PANIC and IRIS2 imaging was done through a $Ks$-band filter as
were the 2MASS images. The differences among $Ks$ filters are much
smaller than the magnitude errors and have therefore been ignored.
As we do not have any colour
information and the effect is very small, we have not converted $Ks$-band magnitudes to $K$-band.
For simplicity, we refer hereafter to the magnitudes as $K$ rather than $Ks$.

Photometry was done in {\sc iraf radprof} using stars from
the 2MASS point source catalogue.
In most fields there were at least four
2MASS stars which were used for calibrating our photometry. 
Because of their smaller fields of view, there were
six PANIC fields which had only two suitable 2MASS stars, two fields
had one and one field had no suitable 2MASS stars. Therefore, all of the
2MASS stars measured in the images were plotted against the
PANIC measured magnitudes, and a best fit line was used to
calibrate the frames that had fewer than four 2MASS stars. Even
in the cases where there were four 2MASS stars to calibrate
an individual field, the photometry agreed, within errors, with
the best fit line for all stars across all fields. In each case, the
resultant errors were carefully assessed based on the photometry
fitting errors and the offset errors from fitting 2MASS
stars.

Several of the PANIC fields were affected either by some cloud, or by
reflections of moonlight onto the detector leading to uneven backgrounds. Stray
light on one quadrant of the IRIS2 chip gave background variations
in some IRIS2 frames. In these
cases the photometric errors are substantially higher.

An
airmass correction was not applied as all images had an airmass $<2$ and the
correction was small compared to the errors.
No Galactic extinction
correction was applied as fewer than 30 of our sources have $|b|<20^{\circ}$.

All of the identifications were compact at $K$-band, such that a 4-arcsec-diameter aperture was sufficient to contain each source without
contamination from nearby objects.

\section{RESULTS}
\label{results}

Fig.~\ref{overlays} shows the ATCA 2368-MHz contours from \citetalias{Bro07},
overlaid on the $K$-band images, for a selection of our sources,
including those discussed individually 
in Section~\ref{notes} and in \citetalias{Bry08}.
Overlay plots for the remainder of our sources can be found in the Appendix.

\begin{figure*}
\begin{center}
\psfig{file=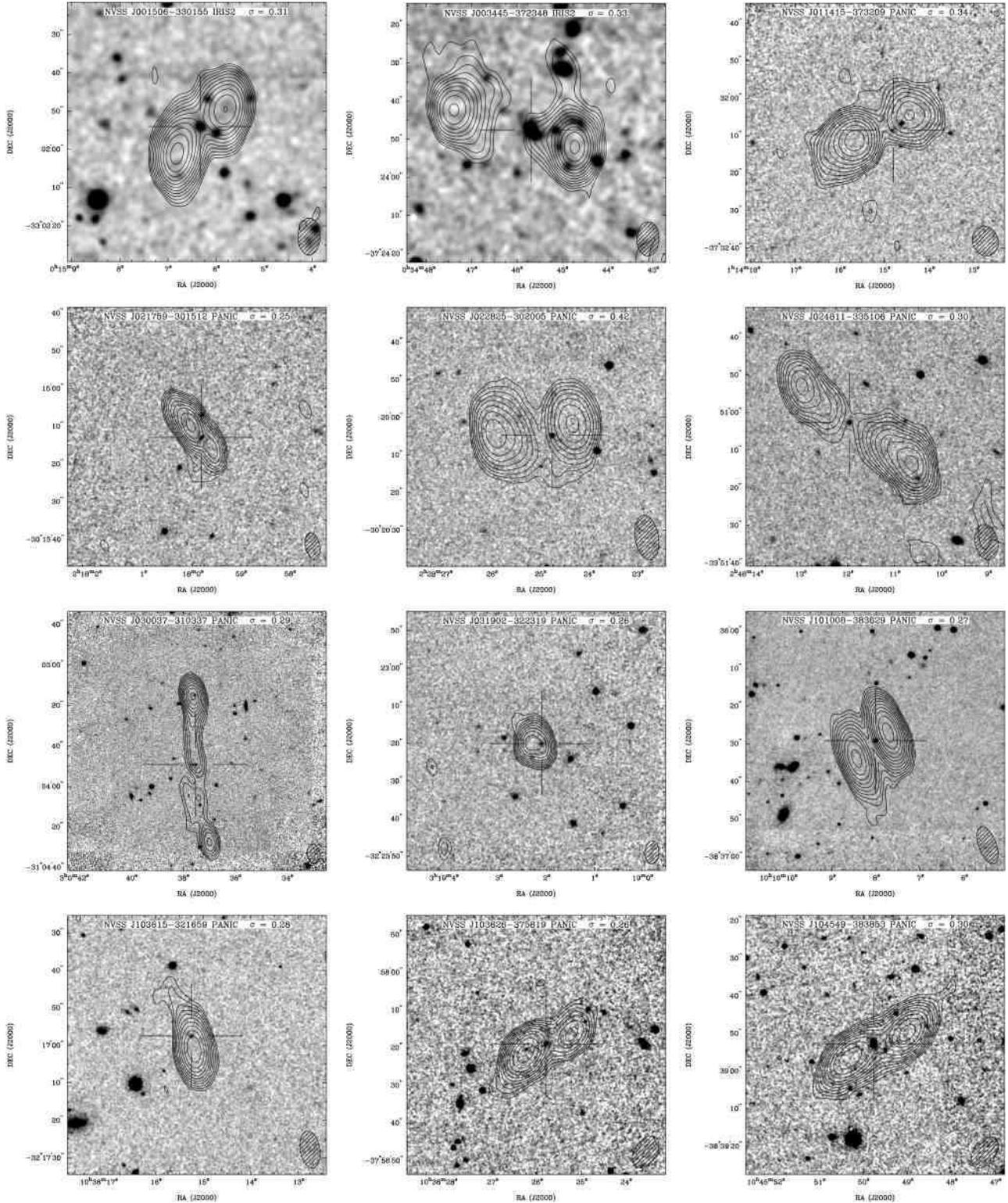, width=17cm}
\caption{2368-MHz ATCA contours overlaid on the $K$-band images for a selection
of our sample, including 
the sources discussed in Section~\ref{notes} and those with both PANIC or IRIS2 images and spectra which will be discussed further in \citetalias{Bry08}.
Overlay plots for the remainder of the $K$-band sample can be found in the Appendix. 
The $K$-band images show in the header which instrument they are from.
They have been smoothed using a gaussian kernel of 3 pixels FWHM for the 
PANIC and IRIS2 images, and 2 pixels FWHM for the 2MASS images.
All the radio contours are from natural-weighted images except 
NVSS~J021759$-$301512, NVSS~J031902$-$322319 and NVSS~J142320$-$365027
which have uniform weighting and slightly higher resolution. The lowest contour is 3 sigma, and the contours
are a geometric progression in $\sqrt 2$. The rms noise ($\sigma$) is shown in the header of each
image in mJy\,beam$^{-1}$.
Crosshairs mark the $K$-band counterpart to the radio source.
The ATCA synthesized beam is shown in the bottom right-hand corner of each panel.}
\label{overlays}
\end{center}
\end{figure*}
\setcounter{figure}{1}
\begin{figure*}
\begin{center}
\psfig{file=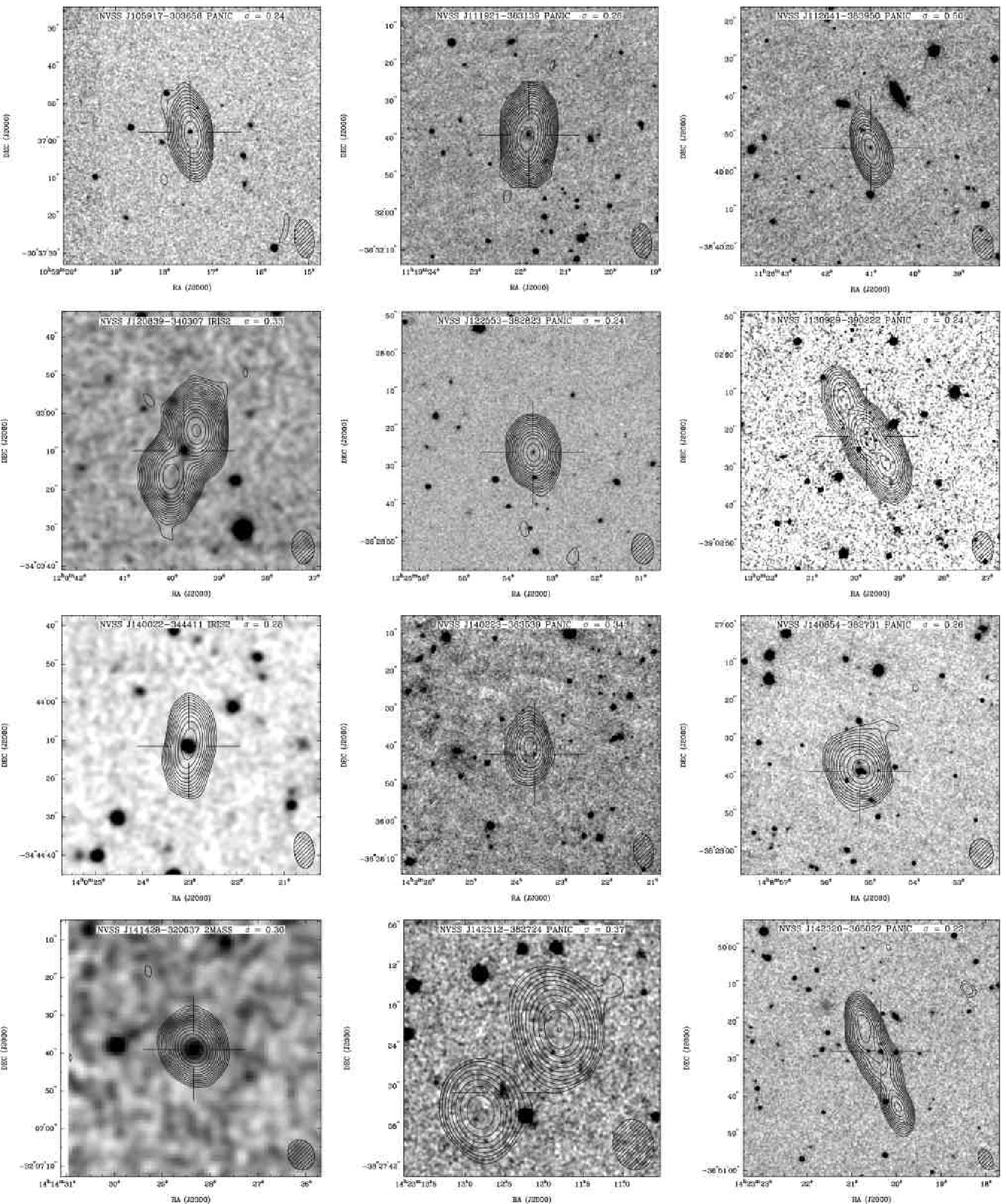, width=17.5cm}
\caption{{\it continued.}}
\end{center}
\end{figure*}
\setcounter{figure}{1}
\begin{figure*}
\begin{center}
\psfig{file=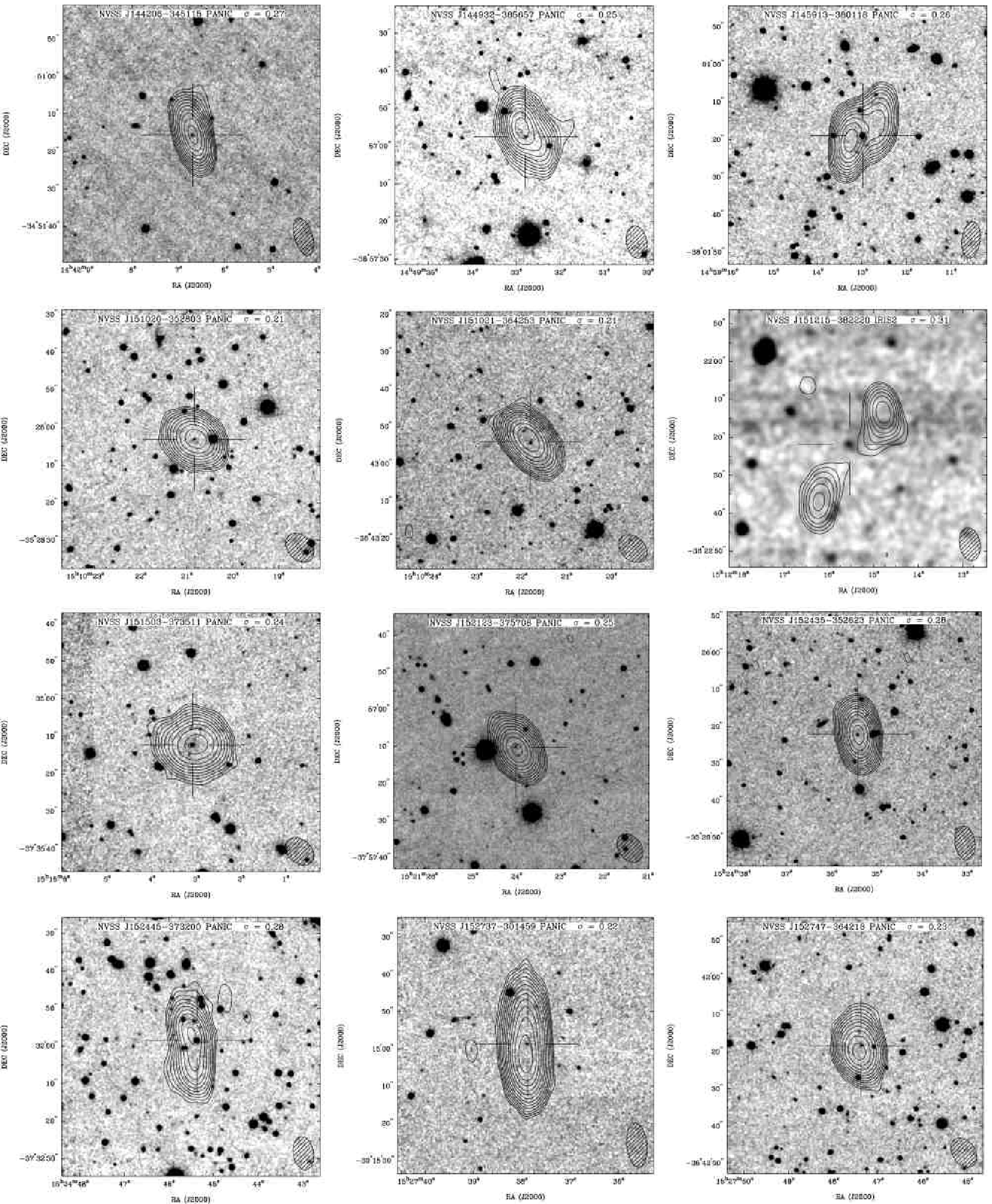, width=17.5cm}
\caption{{\it continued.}}
\end{center}
\end{figure*}
\setcounter{figure}{1}
\begin{figure*}
\begin{center}
\psfig{file=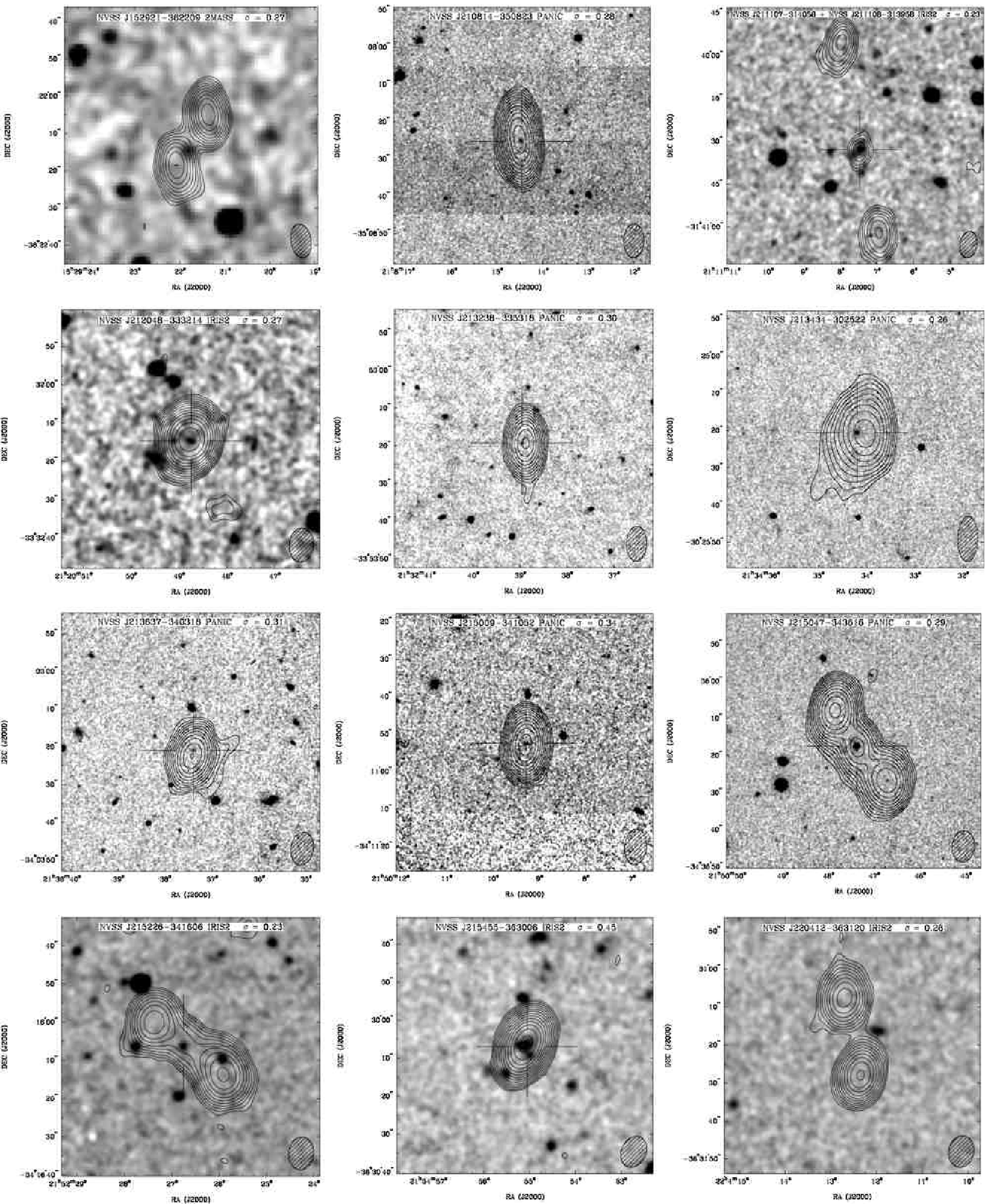, width=17.5cm}
\caption{{\it continued.}}
\end{center}
\end{figure*}
\setcounter{figure}{1}
\begin{figure*}
\begin{center}
\psfig{file=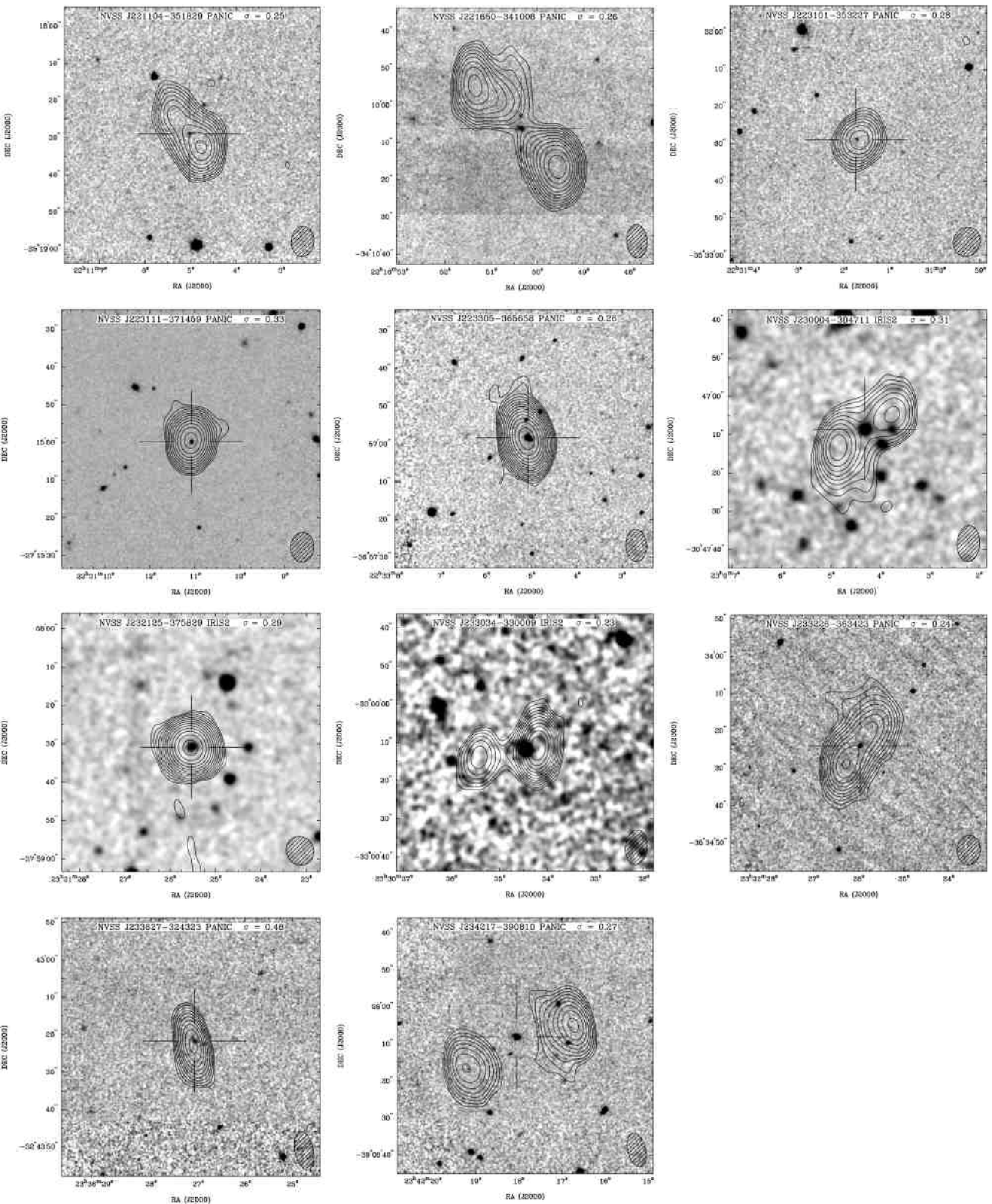, width=17.5cm}
\caption{{\it continued.}}
\end{center}
\end{figure*}

\begin{figure*}
\begin{center}
\psfig{file=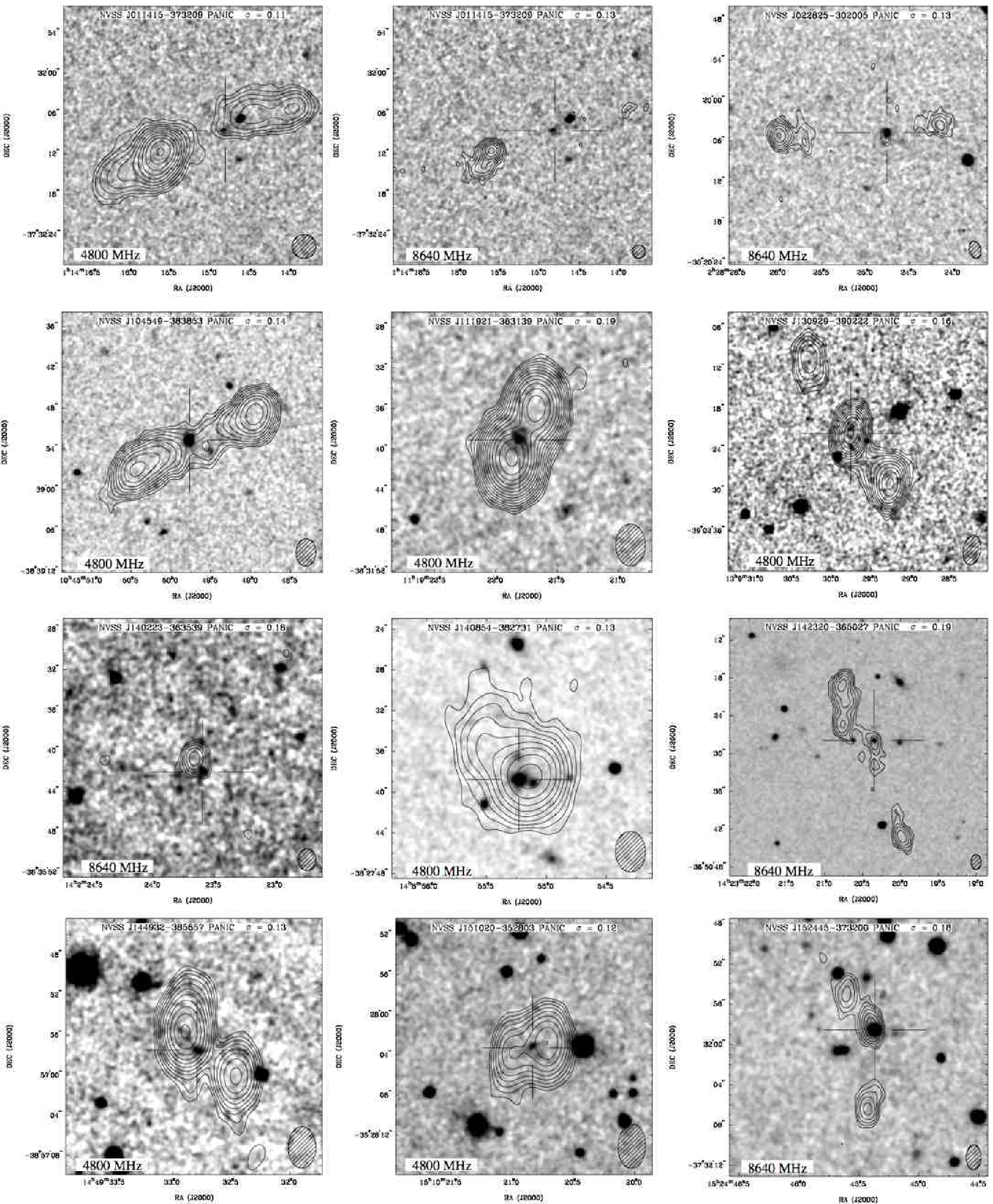, width=17cm}
\caption{4800 or 8640-MHz ATCA contours overlaid on the $K$-band images including 
some of the sources
which had uncertain identifications at 2368\,MHz.
These are discussed in Section~\ref{notes}.
The $K$-band images show which instrument they are from, in the header,
and they have been smoothed using a gaussian kernel of 3 pixels FWHM.
All the radio contours are from natural-weighted images with the
lowest contour at 3 sigma, and the contours
are a geometric progression in $\sqrt 2$. The rms noise ($\sigma$) is shown in the header of each
image in mJy\,beam$^{-1}$.
Crosshairs mark the $K$-band counterpart to the radio source.
The ATCA synthesized beam is shown in the bottom right-hand corner of each panel.}
\label{highRes}
\end{center}
\end{figure*}
\setcounter{figure}{2}
\begin{figure*}
\begin{center}
\psfig{file=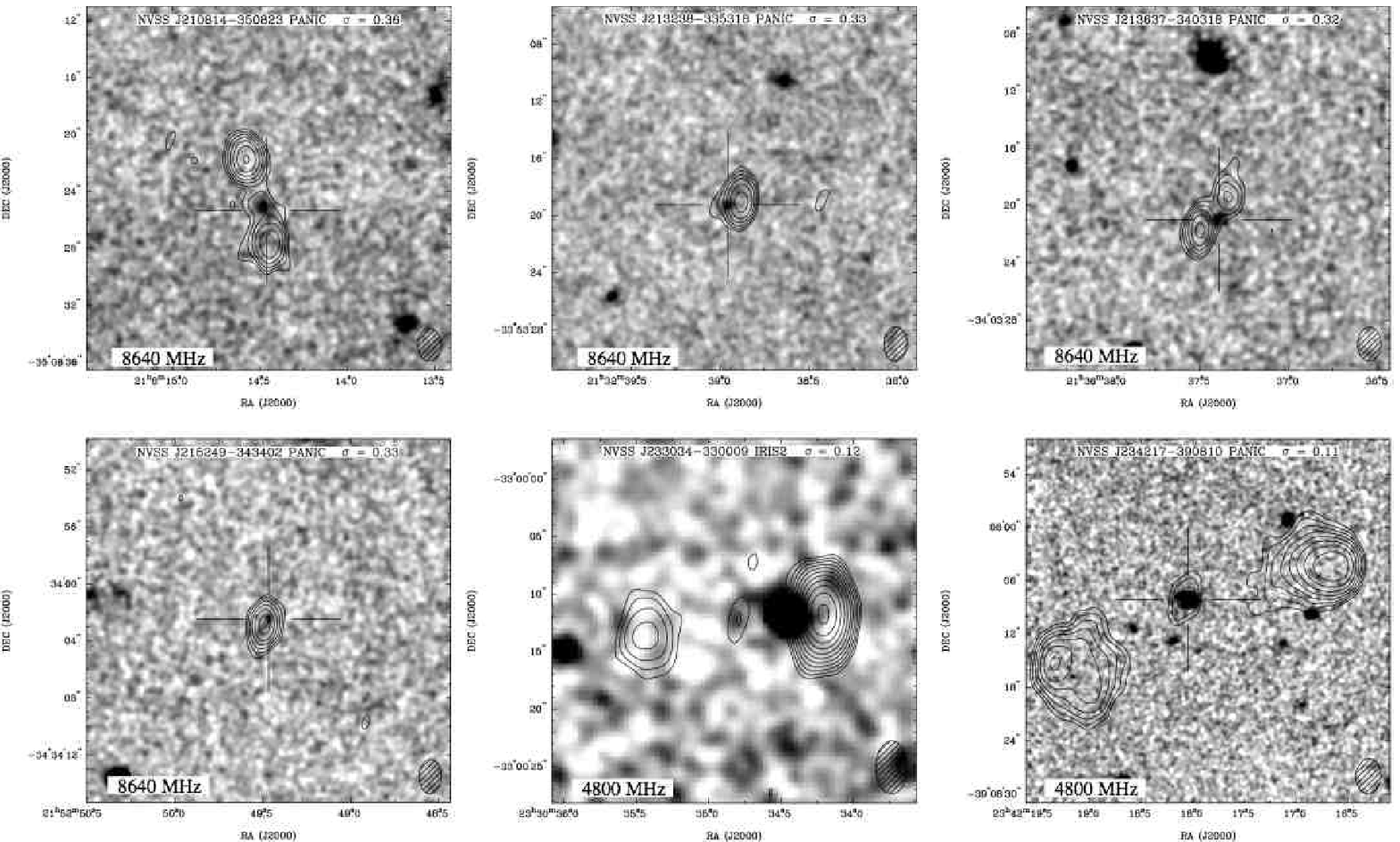, width=17.5cm}
\caption{{\it continued.}}
\end{center}
\end{figure*}

Table~\ref{data} 
lists the data for the objects we observed in $K$-band, or for which
we have an identification which has been followed up with spectroscopy 
(presented in \citetalias{Bry08}).
Some sources not only met our selection 
criteria, but were also part of the SUMSS--NVSS sample 
(\citeauthor{deB04} \citeyear{deB04}; Paper\,I)
and were therefore not reobserved. For those sources, the $K$-band magnitudes and 
positions given are from \citet{deB04}.

The columns in Table~\ref{data} 
are as follows:

\noindent (1) NVSS source name. 

\noindent (2) and (3) $K$-band magnitude in 4-arcsec- and 8-arcsec-diameter apertures
respectively. For sources marked SC, no $K$-band image was obtained as a faint 
SuperCOSMOS identification was found in our overlay plots following the ATCA 
imaging.
``---" means that a $K$ magnitude could not be measured in an 8-arcsec aperture 
because of a nearby source or because of the sky brightness surrounding very faint sources.

\noindent (4) and (5) RA and DEC (J2000) of the $K$-band identification. 
Sources that had no $K$-band or SuperCOSMOS identification are marked ``---".

\noindent (6) $K$-band image exposure time in minutes.

\noindent (7) The origin of the $K$-band image. PANIC is on the
Magellan Baade Telescope, and IRIS2 is on the AAT. `DB' refers to sources from 
\citet{deB04} and `SC' is written for sources without $K$, which were identified
on the SuperCOSMOS UKJ and UKR fields. Two objects are from the 2MASS catalogue.

\noindent (8) Spectral index from the 5-point fit from \citetalias{Bro07}. 
Some sources 
have spectral shapes that could not be fitted with a 5-point
linear or quadratic fit and the spectral index between 843 and 1400\,MHz is shown instead.
For the sources from \citet{deB04}, the spectral index has been 
recalculated using updated flux densities from the latest 
SUMSS catalogue release (version 2.1).
Sources with flattening radio spectra at high frequencies (marked {\it f}) 
have a spectral index fitted at 1400MHz \citepalias[see][]{Bro07}. 
The spectral 
indices 
from \citet{kla06}, used for the sources marked {\it g}, are the 
result of fits to flux densities at 
up to seven frequencies between 0.843--18\,GHz.

\noindent (9) Largest angular size (LAS) from the natural-weighted 2368-MHz images
in \citetalias{Bro07}, given in arcsec. The sources which are also in the SUMSS--NVSS sample
have LAS values from \citet{deB04} or \citet{kla06} and are not
all measured at 2368\,MHz. The LAS values for NVSS~J002431$-$303330
and NVSS~J103615$-$321659 have been updated since \citetalias{Bro07}, with corresponding new position angles
of 19 and 17 degrees respectively.

Table~\ref{data} includes all the sources that meet our selection criteria
from \citetalias{Bro07} (tables 1 and 3), and that have $K$-band magnitudes available
from our observations, 2MASS, or \citet{deB04}. 
NVSS~J230527$-$360534 and NVSS~J231727$-$352606  
are in both the SUMSS--NVSS and our MRCR--SUMSS samples and have $K$-band magnitude 
limits listed in \citet{deB04}, but were reobserved to 
refine the magnitude; our new magnitude is
shown in the table. The combined source NVSS~J023601$-$314204/J023605$-$314235 and NVSS~J103441$-$394957 have red identifications in
SuperCOSMOS and were retained in the sample
because the counterpart   
was not located until we had high-resolution ATCA images. These sources have followup spectroscopy presented in \citetalias{Bry08}.

\begin{table*}

\caption{$K$-band photometry results. Details of each column are given in Section 3.}
\label{data}
\scriptsize
\begin{tabular}{lcccccccc}\hline
\hline

(1)       &              (2)  &  (3)   &          (4)    &  (5)     &      (6)  &          (7)            &       (8)       &        (9)        \\
Name	&		$K$ (4$''$ ap.)			&	$K$ (8$''$ ap.)			&	RA$^{K-{\rm band}}_{{\rm J2000}}$	&	DEC$^{K-{\rm band}}_{{\rm J2000}}$	&	$K$ int. 	&  Origin of			&	$\alpha_{{\rm (5-point\,fit)}}$	&	LAS$_{2368}$		\\
 	&		(mag)			&	(mag)			&	(h m s) 	&	($\circ$ $'$  $''$)					& time (min) &	$K$ imaging	&			&	(arcsec)		\\
\hline
NVSS J000231$-$342614	&		$	18.2	\pm	0.2	$	&	$	17.9	\pm	0.2	$	&	00:02:31.50	&	$-$34:26:16.1	&		 15		&	IRIS2			&	$	-1.07	\pm	0.05	$	&	55.0			\\
NVSS J000640$-$345027	&		$	19.4	\pm	0.1	$	&	$	19.3	\pm	0.4	$	&	00:06:40.81	&	$-$34:50:27.1	&		 9		&	PANIC			&	$	-0.82	\pm	0.05	$	&	1.7			\\
NVSS J000717$-$365557	&		$	18.9	\pm	0.2	$	&	$	19.1	\pm	0.3	$	&	00:07:17.30	&	$-$36:55:56.2	&		 9		&	PANIC			&	$	-0.94	\pm	0.05	$	&	8.7		\\
NVSS J000742$-$304325	&		$	18.9	\pm	0.1	$	&	$	18.6	\pm	0.2	$	&	00:07:42.83	&	$-$30:43:24.2	&		 9		&	PANIC			&	$	-1.16	\pm	0.05	$	&	$<2$			\\
NVSS J001210$-$342103	&	$>$	 	20.0		 	 	&		---				&	---	&	---	&		 9	&	PANIC	&			$	-1.08	\pm	0.05	$	&	36.6			\\
NVSS J001215$-$374113	&		$	20.7	\pm	0.2	$	&	$	19.8	\pm	0.5	$	&	00:12:15.85	&	$-$37:41:13.5	& 18			&	PANIC			&	$	-1.30	\pm	0.04	$	&	$<2$			\\
NVSS J001506$-$330155	&		$	17.6	\pm	0.2	$	&	$	17.1	\pm	0.2	$	&	00:15:06.31	&	$-$33:01:54.0	& 15			&	IRIS2		&	$	-1.00	\pm	0.05	$	&	16.4				\\
NVSS J002037$-$355717	&		$	19.4	\pm	0.2	$	&	$	19.2	\pm	0.4	$	&	00:20:37.66	&	$-$35:57:19.1	& 9		&	PANIC			&	$	-1.10	\pm	0.04	$	&	16.2				\\
NVSS J002431$-$303330	&		$	18.8	\pm	0.1	$	&	$	18.7	\pm	0.1	$	&	00:24:31.80	&	$-$30:33:28.1	& 15		&	IRIS2			&		$-0.87^{\it f}$			&	7.7			\\
\vspace{3mm}
NVSS J002805$-$362049	&		$	18.7	\pm	0.3	$	&	$	18.5	\pm	0.3	$	&	00:28:05.83	&	$-$36:20:50.2	& 15		&	IRIS2			&	$	-0.98	\pm	0.05	$	&	25.9			\\
NVSS J002913$-$353637	&		$	18.1	\pm	0.1	$	&	$	17.5	\pm	0.1	$	&	00:29:13.55	&	$-$35:36:39.4	& 15		&	IRIS2			&	$	-0.92	\pm	0.05	$	&	7.7			\\
NVSS J002915$-$304034	&		$	18.4	\pm	0.1	$	&	$	18.6	\pm	0.3	$	&	00:29:15.30	&	$-$30:40:35.4	& 9		&	PANIC			&	$	-0.70	\pm	0.10$\rlap{$^{\it j}$}	&	23.9			\\
NVSS J003445$-$372348	&		$	17.0	\pm	0.1	$	&	$	16.4	\pm	0.1	$	&	00:34:45.66	&	$-$37:23:47.4	& 15			&	IRIS2		&	$	-1.18	\pm	0.04	$	&	32.7			\\
NVSS J003758$-$340630	&		$	17.7	\pm	0.1	$	&	$	17.2	\pm	0.1	$	&	00:37:58.51	&	$-$34:06:30.8	& 15		&	IRIS2			&	$	-1.08	\pm	0.04	$	&	19.4			\\
NVSS J004000$-$303333	&		$	19.5	\pm	0.2	$	&	$	19.8	\pm	0.4	$	&	00:40:00.01	&	$-$30:33:32.7	& 18		&	PANIC			&	$	-1.17	\pm	0.05	$	&	7.2			\\
NVSS J004030$-$304753	&		$	18.8	\pm	0.2	$	&	$	19.0	\pm	0.4	$	&	00:40:30.68	&	$-$30:47:55.7	& 9		&	PANIC			&	$	-1.26	\pm	0.06	$	&	24.5			\\
NVSS J004136$-$345046	&		$	19.2	\pm	0.3	$	&	$	19.1	\pm	0.6	$	&	00:41:36.22	&	$-$34:50:46.8	& 9		&	PANIC			&	$	-1.11	\pm	0.05	$	&	7.9			\\
NVSS J004147$-$303658	&		$	19.6	\pm	0.3	$	&	$	17.7	\pm	0.3	$	&	00:41:47.59	&	$-$30:36:56.1	& 5		&	PANIC			&	$	-0.95	\pm	0.05	$	&	15.8			\\
NVSS J004230$-$343733	&		$	17.0	\pm	0.1	$	&	$	16.9	\pm	0.1	$	&	00:42:30.38	&	$-$34:37:33.2	& 9		&	PANIC			&	$	-1.02	\pm	0.05	$	&	31.3			\\
\vspace{3mm}
NVSS J004609$-$341049	&		$	17.1	\pm	0.1	$	&	$	17.2	\pm	0.1	$	&	00:46:09.40	&	$-$34:10:49.4	& 9		&	PANIC			&	$	-0.91	\pm	0.05	$	&	29.2			\\
NVSS J005320$-$322756	&	$>$	 	20.4		 	 	&		---		 		&	---	&	---	&	 15	&	PANIC	&			$	-1.08	\pm	0.05	$	&	11.8		\\
NVSS J005402$-$361929	&		$	20.6	\pm	0.5	$	&		---		 		&	00:54:02.56	&	$-$36:19:30.5	&	 18			&	PANIC			&	$	-1.14	\pm	0.05	$	&	2.9		\\
NVSS J005509$-$361736	&		$	20.6	\pm	0.5	$	&		---		 		&	00:55:09.74	&	$-$36:17:36.4	&	 27			&	PANIC			&	$	-0.81	\pm	0.08$\rlap{$^{\it j}$}	&	2.8			\\
NVSS J005716$-$373238	&		$	17.8	\pm	0.1	$	&	$	17.5	\pm	0.1	$	&	00:57:16.54	&	$-$37:32:41.6	&	 9			&	PANIC			&	$	-1.26	\pm	0.05	$	&	11.7			\\
NVSS J010021$-$394347	&		$	18.1	\pm	0.1	$	&	$	18.4	\pm	0.3	$	&	01:00:21.18	&	$-$39:43:46.2	&	 9			&	PANIC			&	$	-0.91	\pm	0.05	$	&	6.1			\\
NVSS J010942$-$341154	&		$	18.7	\pm	0.2	$	&	$	18.1	\pm	0.2	$	&	01:09:42.46	&	$-$34:11:54.5	&	 9			&	PANIC			&	$	-1.14	\pm	0.05	$	&	$<2$			\\
NVSS J011415$-$373209	&		$	19.1	\pm	0.2	$	&	$	17.9	\pm	0.1	$	&	01:14:14.81	&	$-$37:32:08.6	&	 9			&	PANIC			&	$	-1.25	\pm	0.04	$	&	17.1			\\
NVSS J011606$-$331241$^{\it i}$	&		$	18.6	\pm	0.2	$	&	$	18.1	\pm	0.2	$	&	01:16:06.77	&	$-$33:12:42.8	&  &	DB	&			$-1.26^{\it f}$			&	6.4		\\
NVSS J011745$-$352644	&		$	18.6	\pm	0.1	$	&	$	16.6	\pm	0.1	$	&	01:17:45.82	&	$-$35:26:44.5	&	 9			&	PANIC			&	$	-0.94	\pm	0.05	$	&	4.6		\\
\vspace{3mm}
NVSS J012754$-$372236	&		$	18.5	\pm	0.2	$	&	$	18.2	\pm	0.3	$	&	01:27:54.94	&	$-$37:22:34.6	&	 3			&	PANIC			&	$	-1.26	\pm	0.05	$	&	4.9			\\
NVSS J012932$-$385433	&		$	18.5	\pm	0.1	$	&	$	18.3	\pm	0.2	$	&	01:29:32.92	&	$-$38:54:34.5	&	 9			&	PANIC			&	$	-1.02	\pm	0.05	$	&	1.4			\\
NVSS J013314$-$304401	&		$	20.1	\pm	0.5	$	&		---		 		&	01:33:14.99	&	$-$30:44:02.4	&	 9			&	PANIC			&	$	-1.09	\pm	0.06	$	&	$<2$			\\
NVSS J013537$-$364411	&		$	18.5	\pm	0.1	$	&	$	17.7	\pm	0.1	$	&	01:35:37.73	&	$-$36:44:11.8	&	 9			&	PANIC			&	$	-1.35	\pm	0.04	$	&	4.1			\\
NVSS J013554$-$332513	&		$	20.4	\pm	0.6	$	&		---		 		&	01:35:54.02	&	$-$33:25:14.8	&	 9			&	PANIC			&	$	-1.09	\pm	0.04	$	&	11.9			\\
NVSS J014445$-$361945	&		$	18.8	\pm	0.5	$	&	$	19.3	\pm	0.6	$	&	01:44:45.40	&	$-$36:19:45.2	&	 6			&	PANIC			&	$	-1.19	\pm	0.05	$	&	7.0			\\
NVSS J014813$-$384340	&		$	18.4	\pm	0.1	$	&	$	18.8	\pm	0.3	$	&	01:48:13.72	&	$-$38:43:41.7	&	 9			&	PANIC			&	$	-1.20	\pm	0.05	$	&	4.3			\\
NVSS J020201$-$374750	&		$	19.3	\pm	0.2	$	&	$	18.9	\pm	0.3	$	&	02:02:01.57	&	$-$37:47:50.1	&	 18			&	PANIC			&	$	-1.15	\pm	0.05	$	&	17.3			\\
NVSS J021208$-$343111	&		$	19.0	\pm	0.2	$	&	$	18.7	\pm	0.3	$	&	02:12:08.67	&	$-$34:31:11.3	&	 9			&	PANIC			&	$	-1.06	\pm	0.05	$	&	7.0			\\
NVSS J021759$-$301512	&		$	18.3	\pm	0.1	$	&	$	18.2	\pm	0.2	$	&	02:17:59.80	&	$-$30:15:12.8	&	 9			&	PANIC			&	$	-1.04	\pm	0.05	$	&	11.7			\\
\vspace{3mm}
NVSS J022825$-$302005	&		$	18.3	\pm	0.1	$	&	$	18.2	\pm	0.2	$	&	02:28:24.76	&	$-$30:20:04.8	&	 9			&	PANIC			&	$	-1.03	\pm	0.05	$	&	19.5			\\
NVSS J023601$-$314204\vline$^{\it a}$	&	 		SC		 		&	---	 	 		& 02:36:04.04	& $-$31:42:23.4\rlap{$^{\it b}$}	&	 &	SC		&	$	-0.35	\pm	0.09$\rlap{$^{\it j}$}		&	58.8		\\
NVSS J023605$-$314235\vline$^{\it a}$	&	& 		 		 		&		 	 	 		&	 	&	 	&					&		 		 		&	 			\\
NVSS J024523$-$325626	&		$	19.6	\pm	0.3	$	&	$	19.5	\pm	0.5	$	&	02:45:23.83	&	$-$32:56:26.2	&	 9			&	PANIC			&	$	-0.66	\pm	0.09$\rlap{$^{\it j}$}	&	$<2$			\\
NVSS J024811$-$335106	&		$	18.7	\pm	0.2	$	&	$	19.7	\pm	0.6	$	&	02:48:11.96	&	$-$33:51:02.6	&	 9			&	PANIC			&	$	-0.80	\pm	0.05	$	&	33.2			\\
NVSS J025737$-$310104	&		$	19.5	\pm	0.2	$	&		---		 		&	02:57:37.81	&	$-$31:01:04.3	&	 9			&	PANIC			&	$	-0.99	\pm	0.05	$	&	4.4			\\
NVSS J030026$-$322103	&		$	17.4	\pm	0.1	$	&	$	17.4	\pm	0.1	$	&	03:00:26.06	&	$-$32:21:01.5	&	 9			&	PANIC			&	$	-0.97	\pm	0.05	$	&	7.0			\\
NVSS J030037$-$310337	&		$	18.2	\pm	0.1	$	&	$	18.0	\pm	0.1	$	&	03:00:37.54	&	$-$31:03:49.2	&	 9			&	PANIC			&	$	-1.04	\pm	0.07	$	&	72.1			\\
NVSS J030431$-$315308	&		$	18.7	\pm	0.2	$	&	$	19.0	\pm	0.3	$	&	03:04:31.91	&	$-$31:53:08.4	&	 9			&	PANIC			&	$	-1.18	\pm	0.05	$	&	2.4			\\
NVSS J031902$-$322319	&		$	19.2	\pm	0.2	$	&		---		 		&	03:19:02.10	&	$-$32:23:20.1	&	 9			&	PANIC			&	$	-0.70 \pm	0.09$\rlap{$^{\it j}$}	&	5.6			\\
\vspace{3mm}
NVSS J032558$-$320154	&	$>$	 	19.9		 	 	&		---		 		&	---	&	---	&		 9	&	PANIC	&			$	-1.34	\pm	0.05	$	&	26.4			\\
NVSS J034027$-$331711	&		$	19.4	\pm	0.2	$	&	$	18.7	\pm	0.3	$	&	03:40:27.22	&	$-$33:17:11.1	&	 9			&	PANIC			&	$	-0.96	\pm	0.05	$	&	22.9			\\
NVSS J100129$-$364834	&		$	17.9	\pm	0.1	$	&	$	18.0	\pm	0.1	$	&	10:01:28.98	&	$-$36:48:34.0	&	 15			&	IRIS2			&	$	-1.16	\pm	0.06	$	&	13.2			\\
NVSS J101008$-$383629	&		$	18.9	\pm	0.2	$	&		---		 		&	10:10:08.02	&	$-$38:36:29.2	&	 9	&	PANIC		&	$	-1.07	\pm	0.04	$	&	10.7		\\
NVSS J101545$-$385816	&		$	18.2	\pm	0.2	$	&	$	17.9	\pm	0.2	$	&	10:15:45.85	&	$-$38:58:17.0	&	 9		&	PANIC			&	$	-1.13	\pm	0.05	$	&	9.4			\\
NVSS J101730$-$384643	&		$	18.0	\pm	0.2	$	&	$	18.0	\pm	0.2	$	&	10:17:30.50	&	$-$38:46:43.8	&	 9		&	PANIC			&	$	-1.28	\pm	0.05	$	&	$<2$			\\
NVSS J102024$-$342257	&		$	18.7	\pm	0.2	$	&	$	18.7	\pm	0.3	$	&	10:20:24.44	&	$-$34:22:57.3	&	 9		&	PANIC			&	$	-1.13	\pm	0.05	$	&	5.5		\\
NVSS J103441$-$394957	&			SC		 		&		---	 	 		&	10:34:42.06	&	$-$39:49:58.0\rlap{$^{\it b}$} 	&		 & 	SC	&		$	-1.05	\pm	0.05	$	&	29.0		\\
NVSS J103615$-$321659	&		$	19.3	\pm	0.2	$	&		---		 		&	10:36:15.26	&	$-$32:16:57.4	&  9	&	PANIC	&		$	-1.22	\pm	0.05	$	&	10.4		\\
NVSS J103626$-$375819	&		$	18.7	\pm	0.2	$	&	$	18.5	\pm	0.2	$	&	10:36:25.79	&	$-$37:58:19.2	& 9		&	PANIC	&	$	-1.17	\pm	0.05	$	&	13.1			\\
\vspace{3mm}
NVSS J104549$-$383853	&		$	17.3	\pm	0.2	$	&	$	17.3	\pm	0.2	$	&	10:45:49.76	&	$-$38:38:52.9	& 9		&	PANIC			&	$	-0.91	\pm	0.04	$	&	16.1			\\
NVSS J105917$-$303658	&		$	18.8	\pm	0.1	$	&		---		 		&	10:59:17.45	&	$-$30:36:57.5	&  9	&	PANIC	&		$	-1.22	\pm	0.05	$	&	2.2		\\
NVSS J110350$-$352643	&		$	19.0	\pm	0.2	$	&	$	18.6	\pm	0.2	$	&	11:03:50.64	&	$-$35:26:43.2	& 15	&	IRIS2	&			$	-1.21	\pm	0.04	$	&	1.6			\\
NVSS J110753$-$394520	&		$	18.5	\pm	0.1	$	&	$	18.6	\pm	0.2	$	&	11:07:53.64	&	$-$39:45:21.0	& 9	&	PANIC	&			$	-0.91 	\pm	0.10$\rlap{$^{\it j}$}	&	6.6			\\
NVSS J110822$-$390601	&	$>$	 	19.8		 	 	&		---		 		&	---	&	---	&		 9		&	PANIC			&	$	-1.15	\pm	0.05	$	&	3.4			\\
NVSS J111026$-$384544	&		$	17.4	\pm	0.1	$	&	$	17.2	\pm	0.1	$	&	11:10:26.40	&	$-$38:45:43.6	& 9		&	PANIC			&	$	-1.01	\pm	0.04	$	&	2.3			\\
NVSS J111546$-$391410	&		$	16.4	\pm	0.2	$	&	$	16.4	\pm	0.2	$	&	11:15:46.47	&	$-$39:14:10.2	& 9		&	PANIC			&	$	-1.08	\pm	0.04	$	&	20.2			\\
NVSS J111921$-$363139	&		$	18.7	\pm	0.1	$	&	$	19.1	\pm	0.2	$	&	11:19:21.81	&	$-$36:31:39.1	&  45	&	PANIC	&		$	-1.41	\pm	0.04	$	&	$<2$		\\
NVSS J112641$-$383950	&		$	20.3	\pm	0.3	$	&		---		 		&	11:26:41.00	&	$-$38:39:53.5	&  36	&	PANIC	&		$	-1.25	\pm	0.04	$	&	$<2$		\\
NVSS J112920$-$314619	&		$	20.5	\pm	0.6	$	&	$	19.9	\pm	0.7	$	&	11:29:20.70	&	$-$31:46:21.1	& 9		&	PANIC			&	$	-0.79	\pm	0.11$\rlap{$^{\it j}$}		&	17.3		\\
NVSS J113002$-$384532	&		$	19.3	\pm	0.1	$	&	$	19.4	\pm	0.2	$	&	11:30:02.79	&	$-$38:45:35.2	& 45		&	PANIC			&	$	-1.20	\pm	0.05	$	&	$<2$			\\
\hline
\end{tabular}
\end{table*}

\setcounter{table}{1}

\begin{table*}

\caption{$K$-band photometry results {\it continued}
}
\scriptsize
\begin{tabular}{lcccccccc}\hline
\hline

(1)       &              (2)  &  (3)   &          (4)    &  (5)     &      (6)  &          (7)            &       (8)       &        (9)     \\
Name	&		$K$ (4$''$ ap.)			&	$K$ (8$''$ ap.)			&	RA$^{K-{\rm band}}_{{\rm J2000}}$	&	DEC$^{K-{\rm band}}_{{\rm J2000}}$	&	 $K$ int.	&  Origin of			&	$\alpha_{{\rm (5-point\,fit)}}$	&	LAS$_{2368}$		\\
 	&		(mag)			&	(mag)			&	(h m s) 	&	($\circ$ $'$  $''$)			& time (min)		&	$K$ imaging	&			&	(arcsec)		\\
\hline
NVSS J113801$-$392257	&		$	18.8	\pm	0.2	$	&	$	18.9	\pm	0.3	$	&	11:38:01.26	&	$-$39:23:02.3	& 9		&	PANIC			&	$	-1.08	\pm	0.04	$	&	20.1			\\
NVSS J115829$-$374410	&		$	21.4	\pm	0.7	$	&		---		 		&	11:58:29.32	&	$-$37:44:10.8	& 54		&	PANIC			&	$	-1.32	\pm	0.05	$	&	1.2			\\
NVSS J115853$-$382921	&		$	18.2	\pm	0.3	$	&	$	18.4	\pm	0.3	$	&	11:58:53.45	&	$-$38:29:21.5	& 9		&	PANIC		 	&	$	-1.29	\pm	0.04	$	&	1.1			\\
NVSS J120839$-$340307	&		$	18.0	\pm	0.1	$	&	$	17.7	\pm	0.1	$	&	12:08:39.75	&	$-$34:03:09.6	& 15	&	IRIS2	&		$	-1.20	\pm	0.04	$	&	13.5		\\
NVSS J122553$-$382823	&		$	19.2	\pm	0.2	$	&		---		 		&	12:25:53.43	&	$-$38:28:26.2	& 9		&	PANIC		&	$	-1.26	\pm	0.04	$	&	1.4			\\
NVSS J123602$-$364522	&		$	18.5	\pm	0.2	$	&	$	18.4	\pm	0.2	$	&	12:36:02.15	&	$-$36:45:21.6	& 15	&	IRIS2	&			$	-1.22	\pm	0.05	$	&	14.7			\\
NVSS J125448$-$391358	&		$	18.9	\pm	0.3	$	&	$	18.7	\pm	0.3	$	&	12:54:48.83	&	$-$39:13:57.9	& 15	&	IRIS2	&			$	-1.15	\pm	0.05	$	&	$<2$			\\
NVSS J130929$-$390222	&		$	19.6	\pm	0.2	$	&	$	18.5	\pm	0.2	$	&	13:09:29.75	&	$-$39:02:21.6	& 9	&	PANIC	&			$	-1.06	\pm	0.04	$	&	20.9			\\
NVSS J131102$-$372516	&		$	18.6	\pm	0.2	$	&	$	18.1	\pm	0.2	$	&	13:11:02.41	&	$-$37:25:16.8	& 15	&	IRIS2	&			$	-1.14	\pm	0.04	$	&	2.8			\\
\vspace{3mm}
NVSS J131202$-$394600	&		$	17.0	\pm	0.1	$	&	$	17.0	\pm	0.1	$	&	13:12:02.58	&	$-$39:45:56.6	& 9	&	PANIC	&			$	-0.79	\pm	0.05	$	&	16.0		\\
NVSS J132551$-$344606	&		$	17.3	\pm	0.1	$	&	$	16.9	\pm	0.1	$	&	13:25:51.96	&	$-$34:46:07.5	& 15	&	IRIS2	&	 		$	-0.55	\pm	0.09$\rlap{$^{\it j}$}	&	6.6			\\
NVSS J134143$-$370612	&		$	19.0	\pm	0.2	$	&	$	18.6	\pm	0.2	$	&	13:41:43.37	&	$-$37:06:12.2	& 15	&	IRIS2	&			$	-1.14	\pm	0.05	$	&	2.7			\\
NVSS J140010$-$374240	&		$	17.0	\pm	0.1	$	&	$	17.0	\pm	0.1	$	&	14:00:10.58	&	$-$37:42:40.1	& 9	&	PANIC	&			$	-0.86	\pm	0.05	$	&	10.3			\\
NVSS J140022$-$344411	&		$	16.3	\pm	0.2	$	&	$	16.2	\pm	0.2	$	&	14:00:23.01	&	$-$34:44:11.3	& 15	&	IRIS2	&		$	-1.02	\pm	0.05	$	&	8.6		\\
NVSS J140037$-$354439	&		$	18.5	\pm	0.1	$	&	$	18.1	\pm	0.2	$	&	14:00:37.76	&	$-$35:44:40.8	& 9	&	PANIC	&			$	-1.36	\pm	0.09$\rlap{$^{\it j}$}	&	2.4			\\
NVSS J140223$-$363539	&		$	19.8	\pm	0.3	$	&	$	19.9	\pm	0.6	$	&	14:02:23.63	&	$-$36:35:42.2	& 27	&	PANIC	&		$	-1.38	\pm	0.05	$	&	3.8		\\
NVSS J140303$-$384625	&		$	19.3	\pm	0.2	$	&		---		 		&	14:03:03.92	&	$-$38:46:23.4	& 9	&	PANIC	&			$	-1.02	\pm	0.05	$	&	18.5		\\
NVSS J140854$-$382731	&		$	16.9	\pm	0.2	$	&	$	16.6	\pm	0.2	$	&	14:08:55.24	&	$-$38:27:38.7	& 9	&	PANIC	&		$	-1.28	\pm	0.05	$	&	5.3	\\
NVSS J141428$-$320637	&		$	15.5	\pm	0.2	$	&	$	15.5	\pm	0.2	$	&	14:14:28.27	&	$-$32:06:38.7\rlap{$^{\it c}$} 	&		&	2MASS	&		$	-1.27	\pm	0.05	$	&	1.6		\\
\vspace{3mm}
NVSS J142013$-$384113	&		$	18.8	\pm	0.2	$	&	$	18.7	\pm	0.2	$	&	14:20:13.85	&	$-$38:41:13.2	&		 15		&	IRIS2			&	$	-1.19	\pm	0.05	$	&	17.2			\\
NVSS J142312$-$382724	&		$	18.8	\pm	0.2	$	&	$	18.8	\pm	0.2	$	&	14:23:12.53	&	$-$38:27:31.0	&		 9		&	PANIC			&	$	-1.03	\pm	0.05	$	&	16.9			\\
NVSS J142320$-$365027	&		$ 18.3\pm 0.1$		 		&	$17.7 \pm 0.1$	 	  		&	14:23:20.34	&	$-$36:50:28.0	&		 9			&	PANIC			&	$	-0.93	\pm	0.05	$	&	18.5			\\
NVSS J143122$-$353300	&		$	20.0	\pm	0.4	$	&		---		 		&	14:31:22.71	&	$-$35:32:55.7	&		 9	&	PANIC	&			$	-1.18	\pm	0.05	$	&	16.0			\\
NVSS J143806$-$395307	&		$	19.2	\pm	0.5	$	&	$	18.6	\pm	0.5	$	&	14:38:06.28	&	$-$39:53:09.4	&		 15	&	IRIS2	&			$	-1.08	\pm	0.05	$	&	3.0			\\
NVSS J144206$-$345115	&	$>$	 	20.1		 	 	&		---		 		&	14:42:06.69	&	$-$34:51:15.6	&	 18	&	PANIC	&		$	-0.94	\pm	0.05	$	&	2.5	\\
NVSS J144245$-$373827	&		$	17.8	\pm	0.1	$	&	$	17.5	\pm	0.1	$	&	14:42:45.49	&	$-$37:38:29.4	&	 9		&	PANIC			&	$	-1.09	\pm	0.04	$	&	16.4		\\
NVSS J144932$-$385657	&		$	19.8	\pm	0.2	$	&		---		 		&	14:49:32.79	&	$-$38:56:57.5	&	 45	&	PANIC	&		$	-1.23	\pm	0.09$\rlap{$^{\it j}$}	&	7.5		\\
NVSS J145045$-$380012	&		$	16.6	\pm	0.1	$	&	$	16.4	\pm	0.1	$	&	14:50:45.04	&	$-$38:00:13.3	&	 9	&	PANIC	&			$	-0.94	\pm	0.05	$	&	10.2			\\
NVSS J145133$-$350202	&		$	19.6	\pm	0.2	$	&	$	20.0	\pm	0.7	$	&	14:51:33.18	&	$-$35:02:03.3	&	 27	&	PANIC	&			$	-0.76	\pm	0.05	$	&	1.0			\\
\vspace{3mm}
NVSS J145312$-$355542	&		$	17.7	\pm	0.1	$	&	$	17.3	\pm	0.1	$	&	14:53:12.56	&	$-$35:55:42.9	&	 9	&	PANIC	&			$	-0.99	\pm	0.05	$	&	9.1			\\
NVSS J145403$-$370423	&		$	19.5	\pm	0.2	$	&		---		 		&	14:54:03.40	&	$-$37:04:26.3	&		 9			&	PANIC			&	$	-1.25	\pm	0.06	$	&	14.8			\\
NVSS J145603$-$370142	&		$	18.0	\pm	0.3	$	&	$	18.0	\pm	0.3	$	&	14:56:03.52	&	$-$37:01:46.9	&		 15			&	IRIS2			&	$	-1.22	\pm	0.05	$	&	24.2			\\
NVSS J145913$-$380118	&		$	18.3	\pm	0.1	$	&	$	18.2	\pm	0.2	$	&	14:59:12.99	&	$-$38:01:19.0	&		 9			&	PANIC		&	$	-1.12	\pm	0.05	$	&	8.3			\\
NVSS J150026$-$393951	&		$	16.8	\pm	0.1	$	&	$	16.5	\pm	0.1	$	&	15:00:26.29	&	$-$39:39:51.9	&		 15	&	IRIS2	&			$	-1.36	\pm	0.04	$	&	18.1			\\
NVSS J150405$-$394733	&		$	18.3	\pm	0.2	$	&	$	17.6	\pm	0.2	$	&	15:04:05.55	&	$-$39:47:36.9	&		 9		&	PANIC			&		$-1.02^{\it f}$		&	20.7			\\
NVSS J151020$-$352803	&		$	20.2	\pm	0.2	$	&	$	20.5	\pm	0.6	$	&	15:10:20.84	&	$-$35:28:03.2	&	 45	&	PANIC	&		$	-1.35	\pm	0.05	$	&	4.2	\\
NVSS J151021$-$364253	&		$	20.1	\pm	0.2	$	&		---		 		&	15:10:21.80	&	$-$36:42:54.1	&	 45			&	PANIC		&	$	-1.09	\pm	0.05	$	&	6.0			\\
NVSS J151215$-$382220	&		$	17.2	\pm	0.1	$	&	$	16.9	\pm	0.1	$	&	15:12:15.51	&	$-$38:22:21.7	&	15			&	IRIS2		&	$	-1.14	\pm	0.06	$	&	28.1			\\
NVSS J151503$-$373511	&		$	17.3	\pm	0.1	$	&	$	17.1	\pm	0.1	$	&	15:15:03.09	&	$-$37:35:12.2	&	 9			&	PANIC		&	$	-1.49	\pm	0.05	$	&	7.8			\\
\vspace{3mm}
NVSS J151610$-$342718	&		$	18.0	\pm	0.3	$	&	$	17.9	\pm	0.3	$	&	15:16:10.45	&	$-$34:27:18.5	&	 9	&	PANIC	&			$	-1.15	\pm	0.05	$	&	11.5			\\
NVSS J152123$-$375708	&		$	19.3	\pm	0.3	$	&	$	19.4	\pm	0.6	$	&	15:21:24.02	&	$-$37:57:10.2	&	 27	&	PANIC	&		$	-0.89	\pm	0.05	$	&	3.6		\\
NVSS J152149$-$390517	&		$	18.6	\pm	0.1	$	&	$	18.4	\pm	0.1	$	&	15:21:49.13	&	$-$39:05:20.5	&	 15	&	IRIS2			&	$	-1.37	\pm	0.04	$	&	11.6			\\
NVSS J152435$-$352623	&		$	20.9	\pm	0.5	$	&		---		 		&	15:24:35.41	&	$-$35:26:22.0	&	 36	&	PANIC		&	$	-1.17	\pm	0.05	$	&	$<2$	\\
NVSS J152445$-$373200	&		$	18.3	\pm	0.1	$	&	$	18.6	\pm	0.2	$	&	15:24:45.37	&	$-$37:31:58.5	&	 9	&	PANIC		&	$	-1.07	\pm	0.05	$	&	13.5		\\
NVSS J152737$-$301459	&		$	17.6	\pm	0.2	$	&		---		 		&	15:27:37.88	&	$-$30:14:58.6	&	 9	&	PANIC		&	$	-1.30	\pm	0.04	$	&	13.3		\\
NVSS J152747$-$364218	&			$19.6 	\pm	0.2$\rlap{$^{\it d}$}	&	$	18.6	\pm	0.2	$	&	15:27:47.38	&	$-$36:42:18.4	&	 27	&	PANIC		&	$	-1.12	\pm	0.04	$	&	$<2$			\\
NVSS J152921$-$362209	&		$	15.7	\pm	0.3$\rlap{$^{\it d,e}$}	&	$	15.7	\pm	0.3	$	&	15:29:21.77	&	$-$36:22:14.6\rlap{$^{\it c}$}	&	 	&	2MASS		&	$	-1.16	\pm	0.05	$	&	16.0			\\
NVSS J153757$-$362659	&		$	18.9	\pm	0.3	$	&		---		 		&	15:37:57.23	&	$-$36:26:58.2	&		 15		&	IRIS2			&	$	-1.01	\pm	0.05	$	&	24.0			\\
NVSS J154513$-$344834	&	$>$	 	20.8		 	 	&		---		 		&	15:45:13.74	&	$-$34:48:34.3	&		 9		&	PANIC			&	$	-1.10	\pm	0.04	$	&	1.3			\\
\vspace{3mm}
NVSS J201943$-$364542	&		$	18.4	\pm	0.3	$	&	$	18.3	\pm	0.3	$	&	20:19:43.54	&	$-$36:45:43.2	&		 15		&	IRIS2			&	$	-1.17	\pm	0.05	$	&	14.7			\\
NVSS J202720$-$341150	&	$>$	 	20.3		 	 	&		---		 		&	20:27:21.00	&	$-$34:11:52.0	&		 18		&	PANIC			&	$	-1.12	\pm	0.04	$	&	22.9			\\
NVSS J202856$-$353709$^{\it i}$	&		$	16.9	\pm	0.1	$\rlap{$^{\it d}$}	&	$	16.6	\pm	0.1	$	&	20:28:56.77	&	$-$35:37:06.0	&		&	DB		&	$	-1.36	\pm	0.10$\rlap{$^{\it j}$}		&	35.8			\\
NVSS J202942$-$341300	&		$	19.3	\pm	0.2	$	&		---		 		&	20:29:42.21	&	$-$34:13:01.3	&		 9		&	PANIC			&	$	-1.20	\pm	0.05	$	&	5.9			\\
NVSS J202945$-$344812$^{\it i}$	&		$	17.6	\pm	0.1	$	&	$	17.3	\pm	0.1	$	&	20:29:45.82	&	$-$34:48:15.5	&		&	DB	&			$-0.94^{\it g}$		&	18.9		\\
NVSS J204526$-$371625	&	$>$	 	20.1		 	 	&		---		 		&	---	&	---	&		 18		&	PANIC	&			$	-1.11	\pm	0.05	$	&	2.2			\\
NVSS J204601$-$335656	&		$	19.7	\pm	0.4	$	&	$	18.5	\pm	0.4	$	&	20:46:01.09	&	$-$33:56:57.1	&	 9			&	PANIC		 	&	$	-1.13	\pm	0.05	$	&	2.2			\\
NVSS J204859$-$393404	&		$	18.9	\pm	0.1	$	&	$	18.0	\pm	0.1	$	&	20:48:59.01	&	$-$39:34:04.2	&	 9			&	PANIC			&	$	-1.40	\pm	0.05	$	&	8.2			\\
NVSS J210626$-$314003	&		$	18.7	\pm	0.2	$	&	$	18.8	\pm	0.4	$	&	21:06:25.90	&	$-$31:40:01.5	&	 9			&	PANIC			&	$	-1.04	\pm	0.05	$	&	24.2			\\
NVSS J210814$-$350823	&		$	19.5	\pm	0.5	$	&		---		 		&	21:08:14.46	&	$-$35:08:25.3	&	 9	&	PANIC	&		$	-1.08	\pm	0.04	$	&	8.1		\\
NVSS J211107$-$314058\vline$^{\it a}$ 	&		$	18.3	\pm	0.2	$	&	$	17.5	\pm	0.2	$	&	21:11:07.50	&	$-$31:40:32.7	&		 15			&	IRIS2	 	&	$	-1.26	\pm	0.07	$	&	67.8		\\
\vspace{3mm}
NVSS J211108$-$313958\vline	&							&						&		&		&							&	&						 		&			\\
NVSS J212048$-$333214	&		$	19.1	\pm	0.1	$	&	$	18.6	\pm	0.1	$	&	21:20:48.76	&	$-$33:32:14.4	&		 15		&	IRIS2	&		$	-1.07	\pm	0.04	$	&	4.5		\\
NVSS J212706$-$340322	&		$	18.1	\pm	0.2	$	&	$	18.2	\pm	0.2	$	&	21:27:06.05	&	$-$34:03:27.1	& 9			&	PANIC	&			$	-0.90	\pm	0.05	$	&	23.1			\\
NVSS J212803$-$313912	&	$>$	 	18.5		 	 	&		---		 		&	---	&	---	&		 15		&	IRIS2	&			$	-1.12	\pm	0.05	$	&	38.3			\\
NVSS J213238$-$335318	&		$	19.8	\pm	0.4	$	&		---		 		&	21:32:38.95	&	$-$33:53:18.9	& 9	&	PANIC	&		$	-1.43	\pm	0.05	$	&	$<2$		\\
NVSS J213434$-$302522	&		$	16.8	\pm	0.2	$	&	$	16.7	\pm	0.2	$	&	21:34:34.20	&	$-$30:25:20.6	& 9	&	PANIC	&		$	-1.12	\pm	0.05	$	&	8.1		\\
NVSS J213637$-$340318	&		$	19.7	\pm	0.2	$	&	$	19.6	\pm	0.3	$	&	21:36:37.39	&	$-$34:03:21.1	& 36	&	PANIC	&		$	-1.36	\pm	0.05	$	&	3.6		\\
NVSS J214114$-$332307	&		$	19.1	\pm	0.3	$	&	$	19.6	\pm	0.3	$	&	21:41:14.04	&	$-$33:23:09.4	& 15		&	IRIS2			&	$	-1.21	\pm	0.04	$	&	38.6			\\
NVSS J215009$-$341052	&		$	19.6	\pm	0.3	$	&		---		 		&	21:50:09.30	&	$-$34:10:52.4	& 18	&	PANIC	&		$	-1.18	\pm	0.05	$	&	1.7		\\
\hline
\end{tabular}
\end{table*}

\setcounter{table}{1}

\begin{table*}
\begin{minipage}{170mm}

\caption{$K$-band photometry results {\it continued}
}
\scriptsize
\begin{tabular}{lcccccccc}\hline
\hline

(1)       &              (2)  &  (3)   &          (4)    &  (5)     &      (6)  &          (7)            &       (8)       &        (9)         \\
Name	&		$K$ (4$''$ ap.)			&	$K$ (8$''$ ap.)			&	RA$^{K-{\rm band}}_{{\rm J2000}}$	&	DEC$^{K-{\rm band}}_{{\rm J2000}}$	& $K$ int.	&  Origin of			&	$\alpha_{{\rm (5-point\,fit)}}$	&	LAS$_{2368}$		\\
 	&		(mag)			&	(mag)			&	(h m s) 	&	($\circ$ $'$  $''$)		& time (min)			&	$K$ imaging	&			&	(arcsec)	\\
\hline
NVSS J215047$-$343616	&		$	17.4	\pm	0.1	$	&	$	17.0	\pm	0.1	$	&	21:50:47.38	&	$-$34:36:17.5	& 9			&	PANIC			&	$	-1.00	\pm	0.05	$	&	23.4			\\
NVSS J215226$-$341606	&		$	18.3	\pm	0.3	$	&	$	18.0	\pm	0.3	$	&	21:52:26.74	&	$-$34:16:06.2	& 15	&	IRIS2	&		$	-0.95	\pm	0.05	$	&	21.4	\\
NVSS J215249$-$343402	&	$>$	 	20.0		 	 	&		---		 		&	21:52:49.47	&	$-$34:34:02.6	& 9	&	PANIC	&		$	-1.46	\pm	0.05	$	&	$<2$			\\
NVSS J215455$-$363006	&		$	17.7	\pm	0.1	$	&	$	16.9	\pm	0.1	$	&	21:54:55.08	&	$-$36:30:06.8	& 15	&	IRIS2	&		$	-1.17	\pm	0.04	$	&	5.2	\\
NVSS J215717$-$313449	&	$>$	 	20.2		 	 	&		---		 		&	---	&	---	&		 18		&	PANIC			&	$	-0.36	\pm	0.10$\rlap{$^{\it j}$}		&	9.7			\\
NVSS J220412$-$363120	&	$>$	 	19.6		 	 	&	$	17.6	\pm	0.1	$	&	---	&	---	&		  15		&	IRIS2			&	$	-1.18	\pm	0.05	$	&	20.7			\\
NVSS J221104$-$351829	&		$	19.0	\pm	0.2	$	&	$	19.1	\pm	0.5	$	&	22:11:05.04	&	$-$35:18:28.8		& 9			&	PANIC		&	$	-1.12	\pm	0.05	$	&	9.4			\\
NVSS J221542$-$361311	&		$	18.6	\pm	0.1	$	&	$	19.1	\pm	0.4	$	&	22:15:42.98	&	$-$36:13:14.3	& 9		&	PANIC	&			$	-0.82	\pm	0.05	$	&	12.2			\\
NVSS J221650$-$341008	&		$	16.4	\pm	0.3	$	&	$	16.2	\pm	0.3	$	&	22:16:50.34	&	$-$34:10:06.1	& 9		&	PANIC	&			$	-1.04	\pm	0.04	$	&	29.5			\\
\vspace{3mm}
NVSS J221940$-$350100	&		$	18.5	\pm	0.3	$	&	$	18.1	\pm	0.3	$	&	22:19:40.22	&	$-$35:01:02.1	&		 15			&	IRIS2			&	$	-0.94	\pm	0.05	$	&	4.1			\\
NVSS J223101$-$353227	&		$	19.5	\pm	0.2	$	&	$	19.6	\pm	0.6	$	&	22:31:01.72	&	$-$35:32:28.8	&	 9	&	PANIC	&		$	-1.83	\pm	0.05	$	&	2.7		\\
NVSS J223111$-$371459	&		$	17.6	\pm	0.1	$	&	$	17.3	\pm	0.1	$	&	22:31:11.09	&	$-$37:14:59.9	&	 9	&	PANIC	&		$	-1.01	\pm	0.05	$	&	2.4		\\
NVSS J223305$-$365658	&		$	16.6	\pm	0.1	$	&	$	16.3	\pm	0.1	$	&	22:33:05.08	&	$-$36:56:58.2	&	 9	&	PANIC	&		$	-0.90	\pm	0.04	$	&	4.4		\\
NVSS J223953$-$344433	&		$	19.6	\pm	0.3	$	&	$	19.4	\pm	0.5	$	&	22:39:53.99	&	$-$34:44:37.2	&	 18		&	PANIC			&	$	-1.20	\pm	0.06	$	&	25.1			\\
NVSS J225225$-$344144	&		$	19.1	\pm	0.3	$	&	$	18.7	\pm	0.4	$	&	22:52:25.15	&	$-$34:41:43.6	&	 9		&	PANIC			&	$	-0.65	\pm	0.10$\rlap{$^{\it j}$}	&	8.0			\\
NVSS J225719$-$343954$^{\it i}$	&		$	16.7	\pm	0.1	$	&	$	16.5	\pm	0.1	$	&	22:57:19.63	&	$-$34:39:54.6	&  	&	DB	&			$-1.77^{\it g}$				&	$<6$		\\
NVSS J230004$-$304711	&		$	17.4	\pm	0.1	$	&	$	17.0	\pm	0.1	$	&	23:00:04.31	&	$-$30:47:08.7	&	  15	&	IRIS2	&		$	-0.80	\pm	0.05	$	&	15.3		\\
NVSS J230226$-$364744	&		$	18.7	\pm	0.2	$	&	$	18.7	\pm	0.2	$	&	23:02:26.62	&	$-$36:47:45.7	&	 18	&	PANIC	&			$	-0.80	\pm	0.12$\rlap{$^{\it j}$}	&	$<2$			\\
NVSS J230527$-$360534$^{\it h}$ 	&	$>$	 	21.2		 	 	&		---		 		&	---	&	---	&	 99		&	PANIC	&	 			$-1.39^{\it g}$				&	$<1$			\\
\vspace{3mm}
NVSS J230846$-$334810$^{\it i}$	&		$	17.0	\pm	0.1	$	&	$	16.6	\pm	0.1	$	&	23:08:46.73	&	$-$33:48:12.4	& 			&	DB	&	 		$	-1.26	\pm	0.10$\rlap{$^{\it j}	$}	&	29.8			\\
NVSS J231723$-$371934	&		$	19.5	\pm	0.2	$	&	$	19.8	\pm	0.6	$	&	23:17:23.70	&	$-$37:19:32.1	&	 9		&	PANIC	&			$	-1.12	\pm	0.05	$	&	16.4			\\
NVSS J231727$-$352606$^{\it h}$	&	$>$	 	20.7		 	 	&		---		 		&	---	&	---	&	 9	&	PANIC	&		$	-1.28	\pm	0.05	$	&	4.0		\\
NVSS J232007$-$302127	&	$>$	 	19.0		 	 	&		---		 		&	---	&	---	&		 6		&	PANIC	&			$	-1.26	\pm	0.04	$	&	6.3			\\
NVSS J232058$-$365157$^{\it i}$	&		$	18.6	\pm	0.2	$	&	$	18.7	\pm	0.4	$	&	23:20:58.28	&	$-$36:51:59.7	&				&	DB		 	&		$-0.97^{\it g}$				&	$<5$		\\
NVSS J232125$-$375829	&		$	17.9	\pm	0.1	$	&	$	17.9	\pm	0.1	$	&	23:21:25.55	&	$-$37:58:30.7	&	 15	&	IRIS2	&		$	-1.01	\pm	0.04	$	&	$<2$		\\
NVSS J233034$-$330009	&		$	17.2	\pm	0.3	$	&	$	17.1	\pm	0.3	$	&	23:30:34.49	&	$-$33:00:11.5\rlap{$^{\it k}$}	&	 15	&	IRIS2	&		$	-1.18	\pm	0.05	$	&	15.3		\\
NVSS J233226$-$363423	&		$	17.9	\pm	0.1	$	&	$	18.0	\pm	0.1	$	&	23:32:25.96	&	$-$36:34:24.1	&	 9	&	PANIC	&		$	-1.06	\pm	0.05	$	&	10.6		\\
NVSS J233535$-$343330	&		$	18.9	\pm	0.2	$	&	$	18.6	\pm	0.2	$	&	23:35:35.51	&	$-$34:33:32.4	&	 15		&	IRIS2			&	$	-1.08	\pm	0.04	$	&	6.8			\\
NVSS J233627$-$324323	&		$	18.5	\pm	0.1	$	&	$	18.2	\pm	0.2	$	&	23:36:27.09	&	$-$32:43:21.8	&	 9			&	PANIC		&	$	-1.46	\pm	0.05	$	&	$<2$			\\
\vspace{3mm}
NVSS J233729$-$355529$^{\it i}$	&		$	19.2	\pm	0.3	$	&	$	17.9	\pm	0.2	$	&	23:37:29.76	&	$-$35:55:29.0		& 			&	DB		 	&	$	-1.53 	\pm	0.09$\rlap{$^{\it j}$}	&	7.4			\\
NVSS J234145$-$350624$^{\it i}$ 	&		$	16.9	\pm	0.1	$	&	$	16.3	\pm	0.1	$	&	23:41:45.85	&	$-$35:06:22.2	&		&	DB	&			$-1.11^{\it g}$				&	$<5$		\\
NVSS J234217$-$390810	&		$ 17.4 \pm 0.1 $	 	 		&	$ 17.3 \pm 0.1$	 	 		&	23:42:18.05	& $-$39:08:08.0	&		 8		&	PANIC	&			$	-1.01	\pm	0.05	$	&	29.2			\\
NVSS J234235$-$384526	&		$	19.0	\pm	0.1	$	&	$	18.8	\pm	0.2	$	&	23:42:35.04	&	$-$38:45:25.0	&		 9		&	PANIC	&			$	-1.36	\pm	0.05	$	&	9.4		\\
NVSS J235148$-$385546	&		$	19.1	\pm	0.2	$	&	$	19.4	\pm	0.5	$	&	23:51:48.33	&	$-$38:55:46.9	&		 9		&	PANIC	&				$-1.19^{\it f}$		&	6.8			\\
NVSS J235638$-$390923	&		$	18.8	\pm	0.1	$	&	$	18.9	\pm	0.3	$	&	23:56:38.66	&	$-$39:09:25.0	&		 9		&	PANIC	&			$	-1.05	\pm	0.05	$	&	19.6			\\
NVSS J235702$-$372108	&		$	19.2	\pm	0.2	$	&	$	18.3	\pm	0.2	$	&	23:57:02.34	&	$-$37:21:11.8	&		 9		&	PANIC	&			$	-1.10	\pm	0.05	$	&	5.7			\\
NVSS J235754$-$390134	&		$	18.8	\pm	0.1	$	&	$	18.8	\pm	0.3	$	&	23:57:54.51	&	$-$39:01:36.2	&		 9		&	PANIC	&			$	-1.41	\pm	0.04	$	&	$<2$			\\
NVSS J235945$-$330354	&		$	19.1	\pm	0.2	$	&	$	18.4	\pm	0.2	$	&	23:59:45.45	&	$-$33:03:56.8	&		 15		&	IRIS2	&			$	-1.35	\pm	0.05	$	&	2.5			\\
\hline
\end{tabular}
\vspace{-2mm}
\footnotetext[1]{Two NVSS sources have been shown to be the components of a single radio source}
\footnotetext[2]{SuperCOSMOS optical position}
\footnotetext[3]{2MASS position}
\footnotetext[4] {Obscured by an M\,star (see \citetalias{Bry08})}
\footnotetext[5] {2MASS $K$ magnitude in 8-arcsec aperture} 
\footnotetext[6] {Fitted spectral index at 1400\,MHz for sources with spectra that flatten at higher frequencies.}
\footnotetext[7] {Spectral indices from \citet{kla06}}
\footnotetext[8] {These sources were observed by \citet{deB04,deB06} in the SUMSS-NVSS sample with no detection in $K$-band. We have reobserved them and improved the magnitude limit for NVSS~J230527$-$360534.}
\footnotetext[9] {These sources were also part of the SUMSS-NVSS sample \citep{deB04,deB06}, and therefore were not reobserved in $K$-band. The $K$-band magnitudes and positions 
given are from \citet{deB04}.}
\footnotetext[10] {$\alpha_{843}^{1400}$ where $\alpha_{5-{\rm point}}$ was not available from \citetalias{Bro07}. For the sources from \citet{deB04}, the spectral index has been recalculated using updated SUMSS flux measurements from the latest release.}
\footnotetext[11] {Position of the object discussed in Section~\ref{notes}, which may be a star.} 

\end{minipage}
\end{table*}

Of the 175 sources with $K$-band and/or spectroscopic (see \citetalias{Bry08}) 
followup, 93 per cent (162) 
had clear $K$-band counterparts to the radio images, 
two had SuperCOSMOS identifications and 11 had counterparts too faint to 
detect or measure. 

\subsection{4800- and 8640-MHz flux densities}
\label{sec_hires}

The results of the 4800- and 8640-MHz observations of 29 sources from the
MRCR--SUMSS sample are given in Table~\ref{highres_tab}. The columns are the
following:
\newline
\newline
(1) Name of the source in the NVSS catalogue in IAU J2000 format.\newline
(2) The 4800- and 8640-MHz integrated flux densities.
In most cases, these were obtained using the {\footnotesize MIRIAD} Gaussian fitting task
{\footnotesize IMFIT}. However, when there was a significant amount of
diffuse extended emission, the flux densities were calculated by summing
the pixel values in a box or polygon and correcting for the background
level. The typical flux density uncertainties are $\sim$4 per cent at 4800\,MHz 
and $\sim$10 per cent at 8640\,MHz, though in a few cases the 8640\,MHz
flux densities are uncertain by up to $\sim$20 per cent due to poor weather conditions. 
\newline
(3) The two-point observed-frame spectral index $\alpha^{8640}_{4800}$. 
\newline
(4) The seven-point observed-frame spectral index based on a linear fit to 
the 408, 843, 1384, 1400 and 2368 MHz flux densities from \citetalias{Bro07}, as well 
as the 4800- and 8640-MHz data points. For those sources that have curved 
spectra that flatten ({\it f}) at high frequency, we use a quadratic fit to all 
seven data points to derive the observed-frame spectral index at 1400 MHz. 
\newline
(5) The largest angular size (LAS). For single-component resolved sources, 
this is the deconvolved major axis of the elliptical Gaussian used to fit 
the source. For unresolved sources, we estimate a LAS upper limit of 1 
arcsec. For multi-component sources, the LAS is the angular separation of 
the most widely-separated components. The LAS measured at 2368\,MHz (see \citetalias{Bro07} 
and Table~\ref{data}) slightly underestimated the size of NVSS~J1524354$-$352623 which has a 
LAS$=2.1$\,arcsec, while NVSS~J111921$-$363139 is found to be 
a 5.4-arcsec asymmetric double at higher resolution. Another case where we find 
an increase in the LAS for a compact source is NVSS~J005509$-$361736; in Fig.~\ref{appendixHR}, 
this source is shown to be a very asymmetric double, with a LAS of 6.7\,arcsec 
(as opposed to 2.8 arcsec in \citetalias{Bro07}).
 \newline
(6) The deconvolved position angle (PA) of the radio structure, measured 
from the north to the east. For multi-component sources, this is the 
orientation of the most widely-separated components used to calculate the 
LAS.\newline

Note that in columns (5) and (6), we use the 8640 MHz images unless source 
components are resolved out at this frequency, in which case we use the 
4800-MHz images.

\begin{table*}
\begin{minipage}{175mm}
\caption{4800- and 8640-MHz ATCA radio properties. The entries are explained in Section~\ref{sec_hires}.}
\label{highres_tab}
\begin{tabular}{crrrrrr}
\hline
\hline
\multicolumn{1}{c}{Source} & \multicolumn{1}{c}{$S_{4800}$} & \multicolumn{1}{c}{$S_{8640}$} & \multicolumn{1}{c}{$\alpha^{8640}_{4800}$} & \multicolumn{1}{c}{$\alpha$} & \multicolumn{1}{c}{LAS} & \multicolumn{1}{c}{PA} \\
& \multicolumn{1}{c}{(mJy)} & \multicolumn{1}{c}{(mJy)} & & \multicolumn{1}{c}{(7-point fit)} & \multicolumn{1}{c}{(arcsec)} & \multicolumn{1}{c}{($\degr$)} \\
\multicolumn{1}{c}{(1)} & \multicolumn{2}{c}{(2)} & \multicolumn{1}{c}{(3)} & \multicolumn{1}{c}{(4)} & \multicolumn{1}{c}{(5)} & \multicolumn{1}{c}{(6)} \\
\hline
NVSS    J001215$-$374113                & $     10.5    \pm     0.4     $ & $   
4.5     \pm     0.4     $ & $   -1.44   \pm     0.16    $ & $   -1.33   \pm     
0.02    $ & $   1.1     $ & $   52    $       \\
NVSS    J005402$-$361929                & $     31.4    \pm     1.1     $ & $   
14.0    \pm     1.3     $ & $   -1.37   \pm     0.17    $ & $   -1.23   \pm     
0.02    $ & $   1.9     $ & $   11    $       \\
NVSS    J005509$-$361736                & $     50.3    \pm     1.8     $ & $   
19.5    \pm     1.8     $ & $   -1.61   \pm     0.17    $ & $   \cdots          
        $ & $   6.7     $ & $   -68   $       \\
NVSS    J011415$-$373209                & $     28.0    \pm     1.1     $ & $   
11.1    \pm     1.3     $ & $   -1.58   \pm     0.21    $ & $   -1.24   \pm     
0.02    $ & $   27.3    $ & $   -71   $       \\
NVSS    J022825$-$302005                & $     24.2    \pm     1.1     $ & $   
10.8    \pm     1.4     $ & $   -1.37   \pm     0.23    $ & $   -1.08   \pm     
0.03    $ & $   23.8    $ & $   -86   $       \\
NVSS    J101008$-$383629                & $     33.3    \pm     1.2     $ & $   
15.4    \pm     1.5    $ & $   -1.31   \pm     0.18    $ & $   -1.09   \pm     
0.02    $ & $   11.6    $ & $   -48   $       \\
NVSS    J103615$-$321659                & $     10.6    \pm     0.4     $ & $   
4.0     \pm     0.4     $ & $   -1.66   \pm     0.18    $ & $   -1.33   \pm     
0.02    $ & $ 10.4^{\rm a}      $ & $ 17      $       \\
NVSS    J104549$-$383853                & $     36.9    \pm     1.5     $ & $   
19.7    \pm     2.2     $ & $   -1.07   \pm     0.20    $ & $   -0.90   \pm     
0.02    $ & $   20.1    $ & $   -66   $       \\
NVSS    J105917$-$303658                & $     15.3    \pm     0.6     $ & $   
8.2    \pm     0.8     $ & $   -1.06   \pm     0.18    $ & $   -1.16  \pm         
0.02     $ & $   1.0     $ & $   27    $       \\
NVSS    J111921$-$363139                & $     92.4   \pm     3.2     $ & $   
33.2    \pm     3.0     $ & $   -1.74   \pm     0.16    $ & $   -1.42   \pm     
0.02    $ & $   5.4     $ & $   -26   $       \\
NVSS    J112641$-$383950                & $     16.3    \pm     0.6     $ & $   
6.6     \pm     0.6     $ & $   -1.54   \pm     0.17    $ & $   -1.34   \pm     
0.02    $ & $   1.1    $ & $   -42   $       \\
NVSS    J115829$-$374410                & $     22.2    \pm     0.8     $ & $   
9.3    \pm     0.9    $ & $   -1.48   \pm     0.18    $ & $   -1.32   \pm     
0.02    $ & $   1.0     $ & $   4    $       \\
NVSS    J130929$-$390222                & $     43.9    \pm     1.6     $ & $   
22.5    \pm     2.0     $ & $   -1.14   \pm     0.16    $ & $   -1.05   \pm     
0.02    $ & $   21.6    $ & $   33    $       \\
NVSS    J140223$-$363539                & $     10.1    \pm     0.5     $ & $   
4.3     \pm     0.5     $ & $   -1.45   \pm     0.21    $ & $   -1.32   \pm     
0.03    $ & $   3.2    $ & $   -48   $       \\
NVSS    J140854$-$382731                & $     20.1    \pm     1.0     $ & $   
6.4     \pm     0.7     $ & $   -1.95   \pm     0.20    $ & $   -1.30   \pm     
0.03    $ & $   5.6     $ & $   55    $       \\
NVSS    J142320$-$365027                & $     61.7    \pm     2.3     $ & $   
30.0    \pm     3.4     $ & $   -1.23   \pm     0.20    $ & $   -0.96   \pm     
0.02    $ & $   25.6    $ & $   22    $       \\
NVSS    J144206$-$345115                & $     15.5    \pm     0.6     $ & $   
9.6     \pm     0.9     $ & $   -0.82   \pm     0.17    $ & $   -0.93   \pm     
0.02    $ & $   0.9     $ & $   1   $       \\
NVSS    J144932$-$385657                & $     21.5    \pm     0.8     $ & $   
11.8    \pm     1.1     $ & $   -1.02   \pm     0.17    $ & $   \cdots          
        $ & $   6.8     $ & $   48   $       \\
NVSS    J151020$-$352803                & $     9.5     \pm     0.4     $ & $   
6.0     \pm     0.6     $ & $   -0.78   \pm     0.18    $ & $   -1.27    \pm       
0.02      $ & $   3.6     $ & $   -75   $       \\
NVSS    J152435$-$352623                & $     24.4    \pm     0.9     $ & $   
15.0    \pm     1.4     $ & $   -0.82   \pm     0.17    $ & $   -1.12           
\: \rm (f)      $ & $   2.1     $ & $   0    $       \\
NVSS    J152445$-$373200                & $     20.1    \pm     0.8     $ & $   
11.2    \pm     1.2     $ & $   -0.99   \pm     0.20    $ & $   -1.01   \pm     
0.02    $ & $   11.2    $ & $   10     $       \\
NVSS    J204601$-$335656                & $     15.3    \pm     0.6     $ & $   
10.5     \pm     2.1     $ & $   -0.64   \pm     0.35    $ & $   -1.11   \pm     
0.03    $ & $   1.6     $ & $   -6    $       \\
NVSS    J210814$-$350823                & $     32.1    \pm     1.2     $ & $   
18.8    \pm     3.8     $ & $   -0.91   \pm     0.35    $ & $   -1.07   \pm     
0.02    $ & $   5.9     $ & $   15    $       \\
NVSS    J213238$-$335318                & $     12.1    \pm     0.5     $ & $   
7.5     \pm     1.5     $ & $   -0.81   \pm     0.35    $ & $   -1.42   \pm     
0.03    $ & $   1.7     $ & $   -75 $       \\
NVSS    J213637$-$340318                & $     15.7    \pm     0.6     $ & $   
8.9     \pm     1.8     $ & $   -0.97   \pm     0.35    $ & $   -1.28   \pm     
0.03    $ & $   3.2     $ & $   -41   $       \\
NVSS    J215249$-$343402                & $     10.6    \pm     0.4     $ & $   
6.5     \pm     1.3     $ & $   -0.83   \pm     0.35    $ & $   -1.41   \pm     
0.03    $ & $   <1     $ & $   \cdots    $       \\
NVSS    J232125$-$375829                & $     43.8    \pm     1.5     $ & $   
29.1    \pm     2.6     $ & $   -0.69   \pm     0.16    $ & $   -0.93           
\: \rm (f)      $ & $   < 1     $ & $   \cdots  $       \\
NVSS    J233034$-$330009                & $     16.6    \pm     0.7     $ & $   
6.8     \pm     0.6     $ & $   -1.52   \pm     0.16    $ & $   -1.24   \pm     
0.03    $ & $   15.3    $ & $   -83   $       \\
NVSS    J234217$-$390810                & $     23.5    \pm     1.0     $ & $   
13.0    \pm     1.6     $ & $   -1.01   \pm     0.22    $ & $   -0.93   \pm     
0.03    $ & $   33.0    $ & $   -71   $       \\

\hline
\vspace{-4mm}
\footnotetext[1]{LAS obtained from 2368 MHz image; see Table~\ref{data}}
\end{tabular}
\end{minipage}
\end{table*}

Some of the sources which had uncertain identifications, and were observed with
higher resolution at 4800 and 8640\,MHz are shown in 
Fig.~\ref{highRes}
and discussed further in Section~\ref{notes}. Overlay plots of the
remaining 4800- and 8640-MHz observations are given in the Appendix (Fig.~\ref{appendixHR}).

\subsection{Notes on individual sources}
\label{notes}

In Section~\ref{distIDs} and Fig.~\ref{distHist} we look in detail
at where the identification of a double-lobed radio source is statistically
expected to be found. We have used that analysis in the following source notes.

\noindent {\bf NVSS~J011415$-$373209:} 
There are two objects which lie close to the axis joining the lobes, and closer
to the fainter lobe than to the brighter. They have $K$-band 
magnitudes of 18.1 and 19.1. 
We have assumed the identification is the fainter object as it 
aligns with the axis of the lobes in the 
8640-MHz image in Fig.~\ref{highRes}. The fainter object is also closer
to the NVSS centroid and the midpoint of the lobes than the brighter
object and is therefore statistically more likely to be the identification 
(see Section~\ref{distIDs}).

\noindent {\bf NVSS~J021759$-$301512:} Several faint galaxies lie near the radio position.
A uniform-weighted image separates the lobes, showing that
the assumed identification is only 0.3\,arcsec from the midpoint of the lobes, 1.3\,arcsec south-east of the NVSS centroid and 0.2\,arcsec from the radio axis.

\noindent {\bf NVSS~J022825$-$302005:} In the 2368-MHz
image in Fig.~\ref{overlays}, the distances of the identification
from the midpoint (4.7\,arcsec) and radio lobe axis (2.2\,arcsec) are large
compared to the distributions in
Fig.~\ref{distHist}. 
In Fig.~\ref{highRes}, however, the 8640 MHz image shows the brighter object coincident
with the radio core and confirms it as the identification. 

\noindent {\bf NVSS~J030037$-$310337:} The radio structure is a wide
separation triple source. There 
is a $K$-band counterpart coincident with the central radio component.

\noindent {\bf NVSS~J031902$-$322319:} The assumed $K$-band identification is 
very faint and diffuse, and is offset 2.7\,arcsec from the centroid. While 
this offset is significant for a compact source, the uniform-weighted image
is elongated slightly towards the identification. Assuming
the identification is correct, we would expect the source to resolve into
a double at higher resolution.

\noindent {\bf NVSS~J103615$-$321659:} 
The chosen identification is 3\,arcsec from the NVSS centroid and the 2368-MHz
image in Fig.~\ref{overlays} shows faint extension to the north-east. Therefore we
expect this to be a double source with one very faint lobe. We also imaged
this source at 4800 and 8640\,MHz but found that any faint lobe
was resolved out at these frequencies. Based on Fig.~\ref{distHist}, a 
3-arcsec offset is only 0.24$\sigma$ from the mean of the distances from 
the NVSS centroid for the sample.

\noindent {\bf NVSS~J103626$-$375819:} 
The identification lies between the lobes and there is a fainter $K$-band 
object coincident with the stronger lobe. 
We assume that the object between 
the lobes is the counterpart since it is only 0.7\,arcsec from the midpoint of
the lobes and lies directly on the radio axis.
The source would have been misidentified with this fainter object based on the 1384-MHz image
alone, as the double structure was not apparent.

\noindent {\bf NVSS~J104549$-$383853:} There are two $K$-band objects
between the lobes, on opposite sides of the radio axis.
The brighter object is only 0.4\,arcsec from the NVSS centroid, 1.4\,arcsec from the midpoint
of the lobes and 0.8\,arcsec off the radio axis. 
Based on its better alignment with 
the high-resolution radio contours
in Fig.~\ref{highRes} we select this brighter source as the identification.

\noindent {\bf NVSS~J111921$-$363139:} The overlay plot in
Fig.~\ref{highRes} shows a broad extension of the $K$-band emission along
the radio axis in both directions. This is discussed further in Section~\ref{pa}.

\noindent {\bf NVSS~J130929$-$390222:} In Figs.~\ref{overlays} and~\ref{highRes}, the 
central component of this triple radio structure is
coincident with a small cluster of four $K$-band features. The $K$ magnitude
was measured by centering on the four objects, all of which lie within the
4-arcsec-diameter aperture. 
The position listed in Table~\ref{data} is the fitted centroid of these four 
possibly associated objects. 

\noindent {\bf NVSS~J140223$-$363539:} 
We believe the identification is the galaxy which is offset
2\,arcsec south-west of the 2368-MHz peak in the direction that the 
8640-MHz contours (see Fig.~\ref{highRes})
are extended. 
A spectrum of this object will be presented in \citetalias{Bry08}.

\noindent {\bf NVSS~J140854$-$382731:} Two $K$-band objects separated by 
1\,arcsec fall near the peak of the radio emission. The 4800-MHz image
in Fig.~\ref{highRes} favours the brighter object as the
identification.

\noindent {\bf NVSS~J142312$-$382724:} The assumed identification is a 
diffuse, possibly multiple,
$K=18.8$ galaxy directly along the axis between the lobes, closer to the
fainter lobe and 4.3\,arcsec from the midpoint. This is one of two radio 
galaxies that are outliers in Fig.~\ref{distHist}(a), each with possible 
identifications that are 
$>2.8\sigma$ from the mean of the distances from the NVSS 
centroid to the $K$ identification. In this case,
in addition to the
claimed identification there is  
another very faint diffuse
object along the axis between the lobes at 14$^{{\rm h}}$23$^{{\rm m}}$12.2$^{{\rm s}}$ $-$38$^{\circ}$27$^{{\rm m}}$26.8$^{{\rm s}}$ (J2000), which is only 2.2\,arcsec from the
NVSS centroid, and 0.27\,arcsec from the midpoint of the lobes. It is therefore
closer to the midpoint position expected for the identification.
The fainter object is too
faint to measure ($K>20$) and 
we have therefore tentatively listed the brighter
object as the identification.

\noindent {\bf NVSS~J142320$-$365027:} The 8640-MHz image
of this source in Fig.~\ref{highRes} reveals a core, confirming the 
identification. A jet extends towards the southern lobe and  
backflow is seen in the
northern lobe, offset from the radio axis. 

\noindent {\bf NVSS~J144932$-$385657:} 
The 4800-MHz image in Fig.~\ref{highRes} shows that the 
identification lies only
0.8\,arcsec from the axis of the double source, and 0.4\,arcsec from the NVSS 
centroid. Moreover, there are two faint $K$-band objects aligned with the 
north-east jet which are discussed further in Section~\ref{pa}.

\noindent {\bf NVSS~J151020$-$352803:}
At 4800\,MHz (Fig.~\ref{highRes}) the radio morphology resolves into a double
source. The $K$-band emission is extended along
the radio axis (see Section~\ref{pa}). 

\noindent {\bf NVSS~J152445$-$373200:} 
Two objects lie near the centre of the radio structure at 2368\,MHz, 
one compact and one diffuse.
The higher-resolution image in
Fig.~\ref{highRes} was required to confirm that 
the correct identification is not the diffuse galaxy
$\sim 1.5$\,arcsec south-east of the core.
Despite
the stellar appearance of the identification,
a colour image generated from deep $J-$ and $K$-band images
reveals this object to be red like  
other galaxies in the field.

\noindent {\bf NVSS~J213238$-$335318:}
The proposed identification is
offset by 0.9\,arcsec from the NVSS centroid but the 
radio source 
appears to be an incipient double at 8640\,MHz (Fig.~\ref{highRes}).

\noindent {\bf NVSS~213434$-$302522:} 
The identification is 2\,arcsec east of the peak in the 2368-MHz
image. The elongated radio morphology suggests that the source would resolve 
into a double at higher resolution.

\begin{figure}
\psfig{file=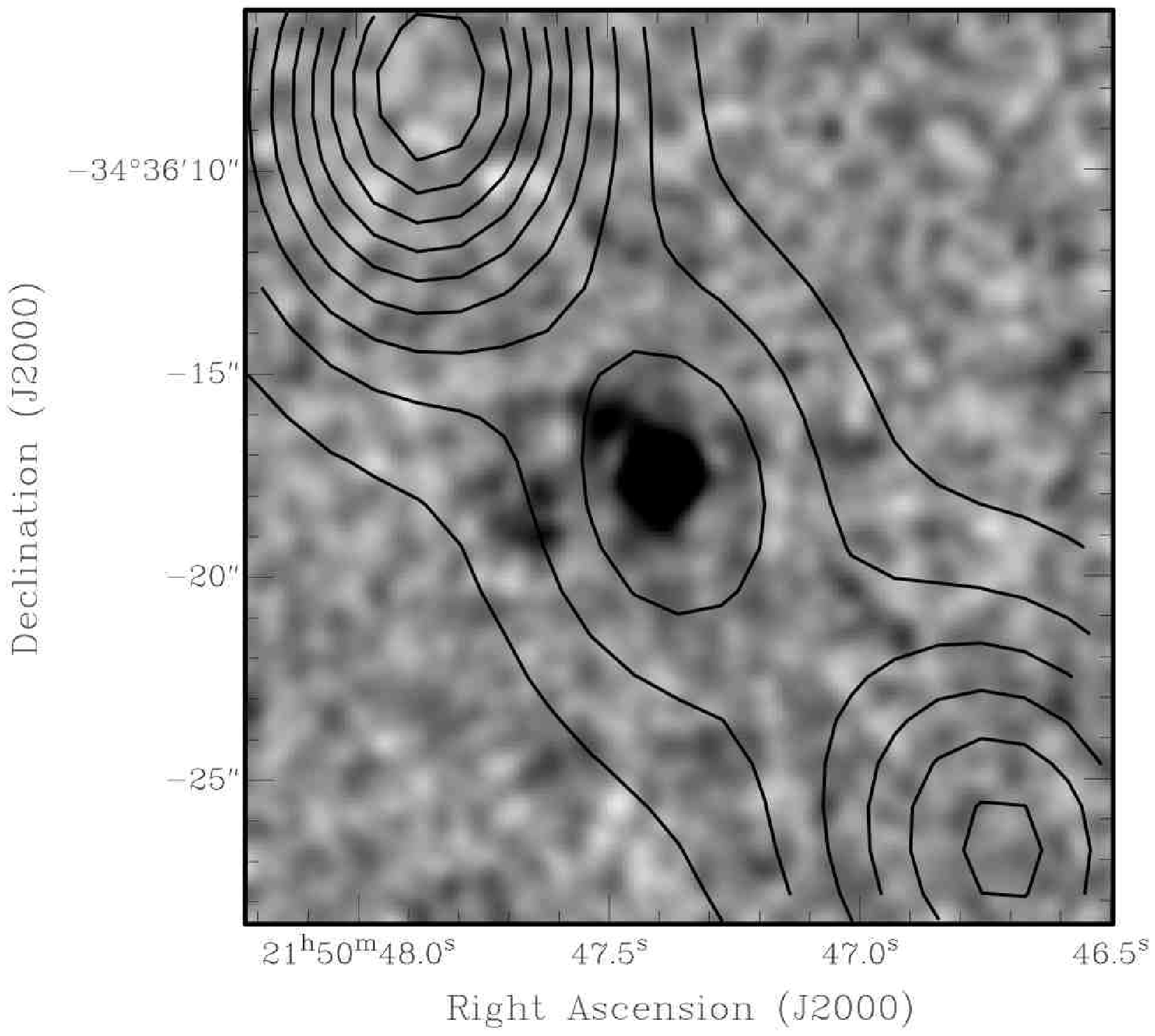, width=7.7cm}
\caption{2368-MHz ATCA contours of J215047$-$343616 overlaid on the
PANIC $K$-band image as in Fig.~\ref{overlays} showing the central 
region where the $K$-band emission is extended along the jet axis.} 
\label{J2150zoom}
\end{figure}

\noindent {\bf NVSS~215047$-$343616:} The $K$-band image in Fig.~\ref{J2150zoom} shows a faint extension $\sim4$-arcsec long, aligned
with the radio jet axis, and directed towards the brighter NE lobe. This
alignment is discussed further in Section~\ref{pa}.

\noindent {\bf NVSS~J220412$-$363120:} Blank field. A $K=18.1\pm0.1$ galaxy lies at 
22$^{{\rm h}}$04$^{{\rm m}}$12.01$^{{\rm s}}$ $-$36$^{\circ}$31$^{{\rm m}}$16.0$^{{\rm s}}$ (J2000),  6.5\,arcsec west of the midpoint of the 
double-lobed radio structure and 7.4\,arcsec from the NVSS centroid.
These are 3.3$\sigma$ and 2.7$\sigma$ respectively from
the mean values in Fig.~\ref{distHist}. 
The $K$-band galaxy itself is $>4.4$-arcsec in extent and 
it contributes to the
8-arcsec-aperture magnitude in Table~\ref{data}. The
accuracy of the astrometry on the IRIS2 and radio images is 
$\pm0.3$\,arcsec and $<1$\,arcsec respectively. 
While we
have treated the radio source as having no $K$-band identification, it is nevertheless possible that this very misaligned
galaxy may be associated.

\noindent {\bf NVSS~J221650$-$341008:} 
The bright diffuse object marked in Fig.~\ref{overlays} is 1.0\,arcsec
from the midpoint of the lobes, and 0.6\,arcsec from the radio
axis and is therefore the assumed identification. 
The fainter object to the north is only 3\,arcsec away and contaminates
the 8-arcsec-aperture magnitude.

\noindent {\bf NVSS~J233034$-$330009:}
On the basis of our original 13- and 20-cm images, we initially chose the 
identification to be the
bright $K$-band object between the lobes and close to the brightest lobe.
However, the higher-resolution image in Fig.~\ref{highRes} has revealed
a weak core which is coincident with a faint peak in the $K$-band image with
$K>19$. 

\noindent {\bf NVSS~J234217$-$390810:} 
The high-resolution 4800-MHz image in Fig.~\ref{highRes} was required to
identify a core coincident with a $K=17.4 \pm 0.1$ galaxy. Both lobes
show backflow towards the core.

\subsubsection{Interlopers in the sample}

While our selection method is necessary to extract the best high-redshift
candidates from large radio catalogues, there will unfortunately always
be interlopers that contaminate the sample. 
In \citetalias{Bro07} we already identified several sources which were excluded
from our sample. Here we discuss another three sources that have been excluded.

\begin{figure*}
\begin{minipage}[]{0.47\textwidth}
\centerline{\psfig{file=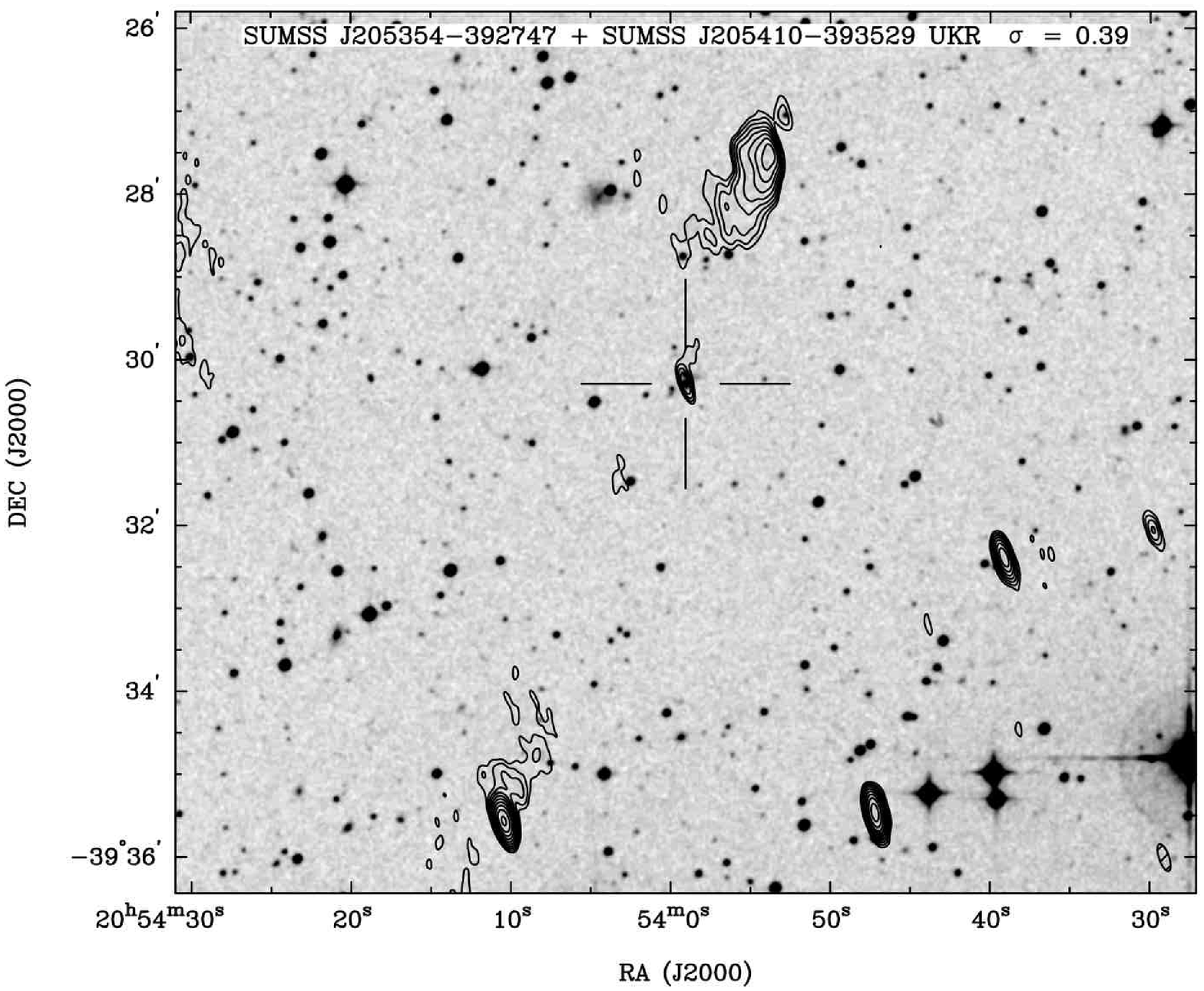,width=7.0cm,angle=-90}}
\end{minipage}%
\hspace*{10mm}
\begin{minipage}[]{0.47\textwidth}
\centerline{\psfig{file=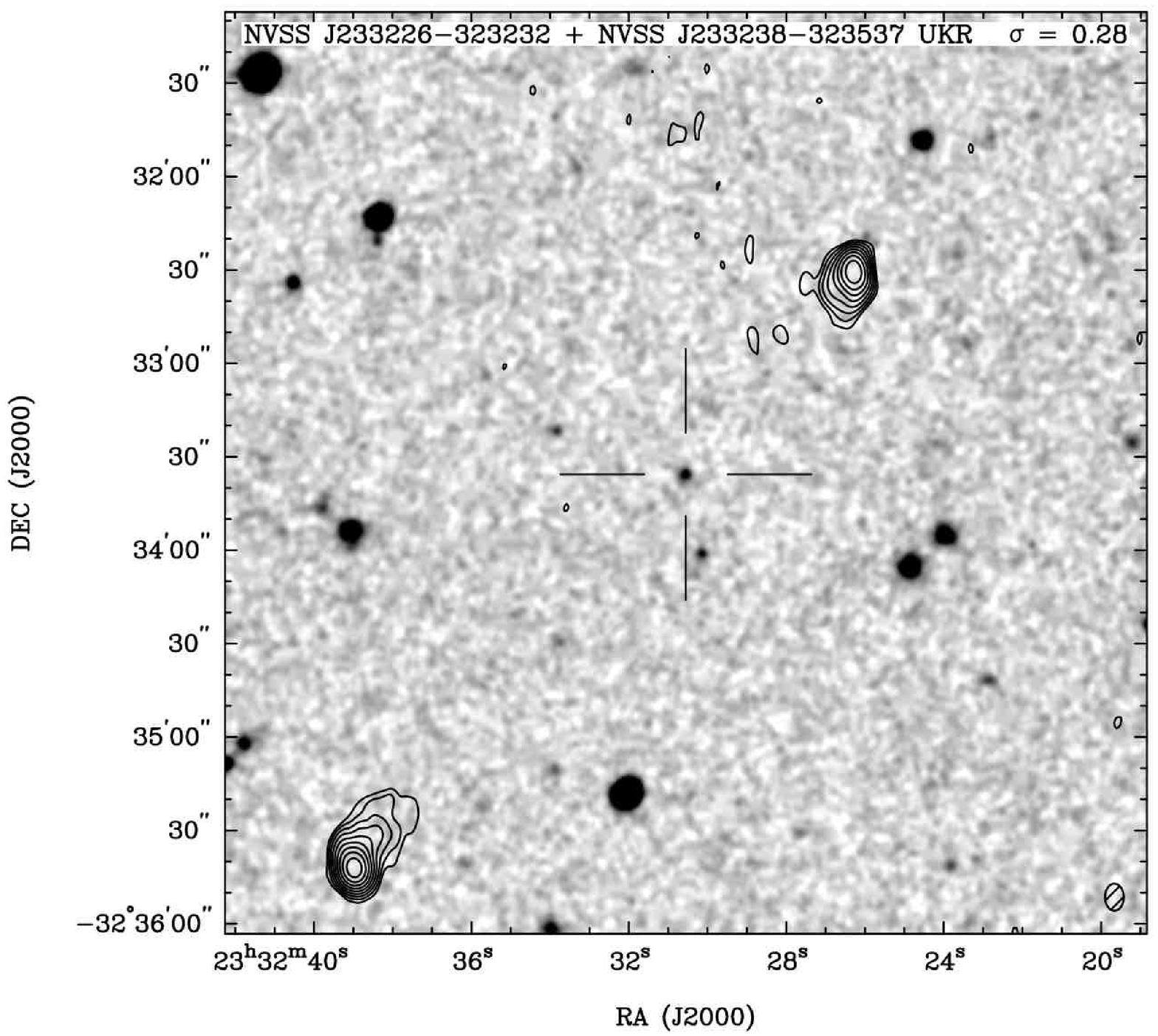,width=6.5cm,angle=-90}}
\end{minipage}
\caption{Smoothed SuperCOSMOS UKR 
images are overlaid with the 1384-MHz (left image) and 2368-MHz (right image)
ATCA contours for two of the sources found to be interlopers in the sample. 
The lowest contour is 3 sigma, and the contours
are a geometric progression in $\sqrt 2$. The rms noise ($\sigma$) is shown in the header of each
image in mJy\,beam$^{-1}$.
The ATCA synthesized beam is shown in the bottom right-hand corner of each panel.
Left: NVSS~J205410$-$393530 was similarly found to be one component of the south-east
lobe of this giant source which has a core, lobes and multiple
hotspots. The identification is a $z=0.167$ QSO, marked by the cross-hairs. 
Right: The source NVSS~J233238$-$323537 in \citetalias{Bro07} had been found to
be the south-east lobe of this giant double source. The north-west lobe
is listed in NVSS as a separate source, NVSS~J233226$-$323232.
The south east lobe is compact and round in the
NVSS image, but the ATCA image reveals low-surface-brightness emission
extended towards the other lobe. The marked object is a star and the host galaxy
has not been identified. }
\label{bigDs}
\end{figure*}

Each of the sources below and in Fig~\ref{bigDs} are in the NVSS catalogue and were listed
in our MRCR--SUMSS sample in \citetalias{Bro07}. However, they have each turned out
to be individual lobes of a giant double or triple source.
Our
selection process is designed to eliminate catalogue sources which are
actually a single lobe of a large double or triple, by
removing sources which have another source within 100\,arcsec.
Then every source
was examined with the NVSS contours overlaid on a 6-arcmin field from SuperCOSMOS.
At this stage there were several sources eliminated as they clearly had
a second lobe within 3\,arcmin. Several other sources had an
elongated morphology characteristic of a lobe, and a second lobe
was then identified in a larger image.
These steps are, in general sufficient to remove most
sources which are lobes of a double.

{\noindent\bf NVSS~J205410$-$393530}: This is 
the lobe of an 8.60-arcmin source which has
a core, lobes and hotspots along the jets.
There is a $K$-band object at 20$^{{\rm h}}$54$^{{\rm m}}$10.67$^{{\rm s}}$ $-$39$^{\circ}$35$^{{\rm m}}$34.0$^{{\rm s}}$ (J2000) that appeared to align with the radio
source, and the magnitude of this object was measured with PANIC to be
$19.5\pm0.4$. Once we discovered NVSS~J205410$-$393530 is in fact a lobe, the core of the giant source
was found to have an identification at 20$^{{\rm h}}$53$^{{\rm m}}$59.07$^{{\rm s}}$ $-$39$^{\circ}$30$^{{\rm m}}$18.1$^{{\rm s}}$ (J2000)
with a 2MASS $K$-band magnitude of $13.58\pm0.05$. A spectrum has identified this object to be
a $z=0.167$ QSO, with a linear size of 1.46\,Mpc and a 1.4-GHz luminosity of 
$1.2\times10^{25}$\,W\,Hz$^{-1}$. The new total flux densities at
843 and 1400\,MHz are $232\pm7$ and $160\pm4$\,mJy respectively, including
all the source components, giving $\alpha_{843}^{1400}=-0.73\pm0.08$. 

{\noindent\bf NVSS~J212553$-$311043}:
The source listed in \citetalias{Bro07} has turned out to be the southern lobe of a triple with a largest
angular size of 3.9\,arcmin. The 
identification is at 
21$^{{\rm h}}$25$^{{\rm m}}$53.72$^{{\rm s}}$ $-$31$^{\circ}$08$^{{\rm m}}$58.9$^{{\rm s}}$ (J2000) 
and has magnitudes of $K=14.6\pm0.1$ (2MASS), $B_{J}=18.7$ and $R=17.9$ (SuperCOSMOS). The total
flux densities at 843 and 1400 MHz are $106\pm6$ and $91\pm3$\,mJy respectively, resulting in a flat spectral
index of $-0.30\pm0.13$.

{\noindent\bf NVSS~J233238$-$323537}: 
This source was missed as a lobe of a double because, firstly, the lobes are
separated by 4.15\,arcmin and, secondly, because the source
from the NVSS catalogue appears to be compact and round with no
extension. 
Once it was imaged in $K$-band with no identification, and the ATCA
images displayed low-surface-brightness characteristic of backflow emission, 
it was clear that we were seeing one lobe of a wide-separation double.
The total NVSS and SUMSS flux densities are $114\pm3$ and 
$160\pm4$\,mJy respectively, giving $\alpha_{843}^{1400}=-0.67\pm0.07$.
There is a bright object along the radio lobe axis
at 
23$^{{\rm h}}$32$^{{\rm m}}$30.57$^{{\rm s}}$ $-$32$^{\circ}$33$^{{\rm m}}$35.77$^{{\rm s}}$ (J2000), 
which is an M star, and the radio galaxy host has not been identified.

\subsubsection{A misidentification}

In \citetalias{Bro07} we suggested that the double-lobed
source, NVSS~J101215$-$394939 had an identification
in SuperCOSMOS. While the listed identification
is along the radio lobe axis, that object has since been shown to
have a stellar spectrum and, therefore, is not the correct 
counterpart. This source will require $K$-band imaging
to find the correct identification.

\section{Discussion}

\subsection{Spectral energy distributions}
\label{SEDs}

In \citetalias{Bro07}, we fitted the spectral energy distributions (SEDs) from
408 to 2368\,MHz. We have now redone these fits with the addition of 
the new 4800- and 8640-MHz data.
Of the 29 sources in Table~\ref{highres_tab}, 25 (86 per 
cent) can be fitted with a single power law between 408 and 8640 MHz in 
the observed frame. Two sources are found to flatten at higher 
frequencies, which may indicate an increased contribution 
from a core and/or hotspots. The remaining two sources (NVSS~J005509$-$361736 
and NVSS~J144932$-$385657) can not be fitted with a 
linear or a quadratic function, as was the case in \citetalias{Bro07} between 408 and 
2368 MHz. 

Our results are consistent with \citet{kla06} who found 89 per cent 
of the 37 SUMSS--NVSS sources were well characterized by a single power law, while the
remainder flattened at higher frequencies.
This is despite the fact 
that \citet{kla06} used matched-resolution ATCA observations between 
2368 and 18\,496 MHz, while our ATCA observations both in \citetalias{Bro07} and this 
paper were all made with a 6 km array configuration. Thus, while some source 
components are clearly resolved out at high frequency (see Figs.~\ref{highRes} 
and~\ref{appendixHR}), the 
effect is not significant enough to alter the shapes of the SEDs.

\subsection{Identification of the host galaxies}
\label{distIDs}

The environments around HzRGs are likely to be denser and more disturbed
than around nearby radio galaxies \citep[e.g.][]{Car97, Pen01, Rot03}. Therefore the expected position of the
optical counterparts relative to the radio emission may not obey the same rules
as found in low-redshift samples such as 3C \citep{Lai83} and the Molonglo 
Southern 4Jy sample \citep[MS4;][]{bur:06}.
In previous HzRG studies, the process of identifying the hosts of USS radio 
galaxies from $K$-band images has been typically based only on which $K$ object
aligns best, by eye, with the radio contours. 
In order to quantify
this process statistically, we have measured the location of the identified 
host galaxy relative
to the features of the radio emission for the double sources in our sample. This information has
then been used to assess
the reliability of identifications for which there are several nearby
$K$-band candidates or where the potential identification is offset from its
expected location; for example, one may tend to look for an identification
closer to the brighter radio lobe. 
Our aim was to assess how far a potential
identification could be displaced from the radio features, and still be
considered statistically likely to be the identification.

\begin{figure*}
\centerline{\psfig{file=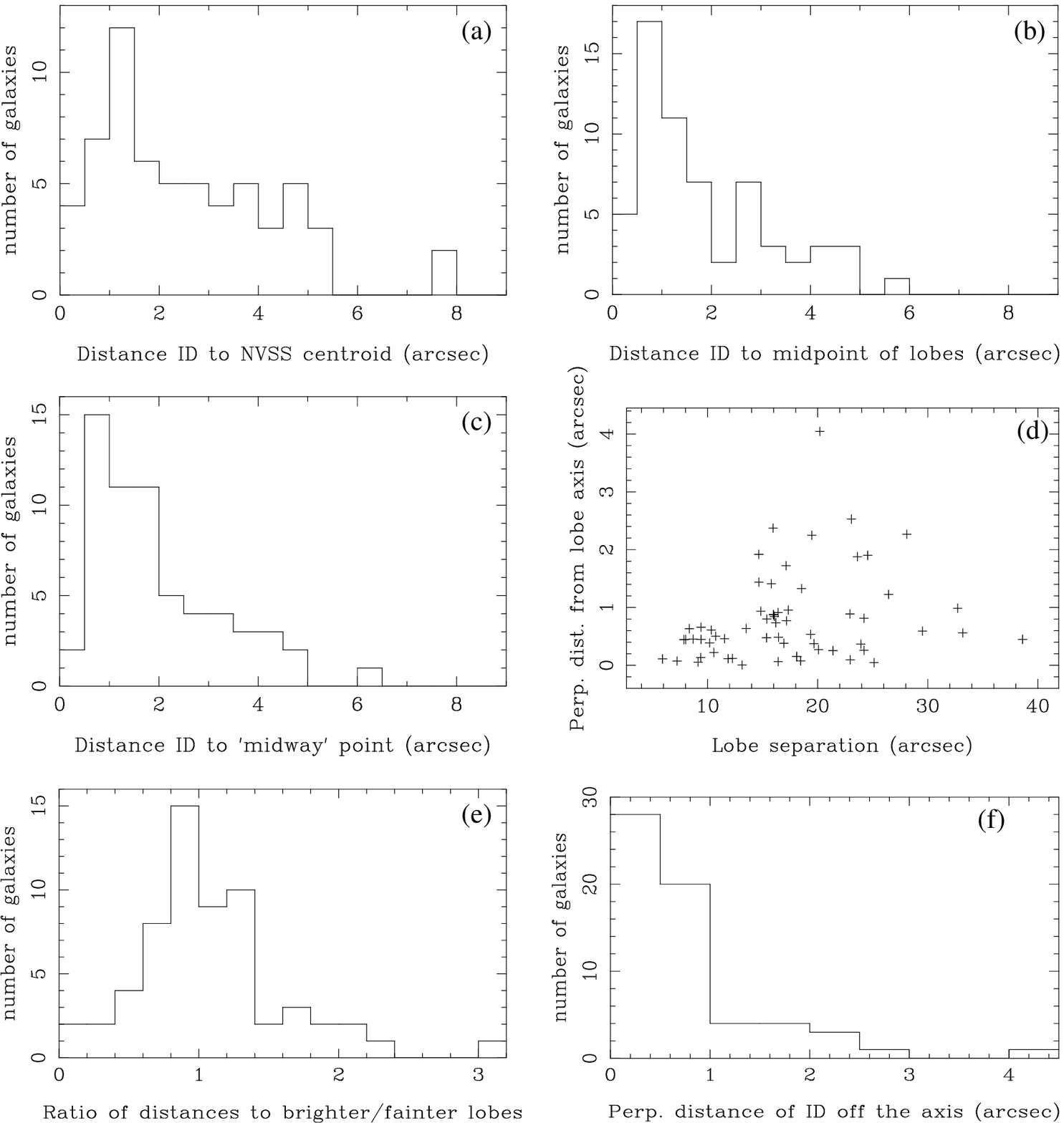,width=14.5cm}}
\caption{Histograms of the angular distances from the proposed $K$-band identification to (a) 
the NVSS centroid, (b) the midpoint of the lobes, and (c) the `midway point' 
(half way between the centroid and the midpoint). (d) 
Distance between the lobes versus the perpendicular distance of the
identification from the axis joining the lobes, for the double sources in
our sample.
(e) Histogram of the ratio of the distance to the brighter lobe to the distance 
to the fainter lobe. (f) Histogram of the perpendicular distance of the 
identification from the axis of the lobes.
}
\label{distHist}
\end{figure*}

The most difficult decisions arise in complex double structures 
rather than compact sources in which the centroid is well defined. In
order to get a statistical picture of the locations of potential 
identifications, angular separations 
were measured from the 
identification to several features of the radio structure for all of the 
double sources in our sample.
The first of these features is the 
radio centroid; since our sample consists mostly of sources $\leq 1$\,arcmin in size,
NVSS is expected to record an accurate 1.4-GHz centroid position.
Next we found the angular distances from the identification
to the midpoint of the lobes, and the `midway point' defined by
\citet{bur:06} 
to be halfway between the midpoint and the centroid.
Lastly, the perpendicular distance of the identification from the
axis joining the lobes was measured.
The resulting distributions
of these distances are shown in Fig.~\ref{distHist}, with the accompanying
statistics in Table~\ref{distHisttable}.

In Table~\ref{distHisttable},
the median, mean and standard deviations are given for the distributions 
of distances to identifications in our sample. 
The midpoint has the lowest median and mean, 
indicating that it is statistically a more likely place
to find the identification compared to the midway point or centroid. 
\citet{bur:06} found the most likely position for the optical
identification for the MS4 was the `midway 
point', based on a smaller median and mean distance. The difference between
both the means and standard deviations for the `midway point' and the midpoint in
our sample is negligible. It is interesting that the centroid position is the least 
likely of the three locations to find the optical counterpart, suggesting that  
low-surface-brightness emission is less important in finding the correct
identification. 66 per cent of optical counterparts lie within 2\,arcsec of the
midpoint of the lobes, while only 48 per cent are as close as this to the centroid.

\begin{table}
\caption{Mean, median and standard deviation for the distributions of
distances from the $K$-band identification to the three radio positions -- 
NVSS centroid, lobe midpoint and 
`midway point' -- in Fig.~\ref{distHist}, given in arcsec. The values
for the distances to the brighter
and fainter lobes have been normalised to the lobe separation.
}
\begin{tabular}{lccc}
\hline
\hline
Radio position & Mean ($''$) & Median ($''$) & $\sigma$ \\
\hline
NVSS centroid & 2.6 & 2.4 & 1.8 \\
Midpoint & 1.9 & 1.4 & 1.4 \\
`Midway point' & 1.9 & 1.6 & 1.3 \\
\hline
\hline
Radio position  &\multicolumn{2}{c}{Normalised} & $\sigma$ \\
 & Mean & Median &  \\
\hline
Brighter lobe & 0.49 & 0.50 & 0.12 \\
Fainter lobe & 0.51 & 0.51 & 0.13 \\
\hline
\label{distHisttable}
\end{tabular}
\end{table}

The statistics for the normalised angular
distances to the brighter and fainter lobes are indistinguishable.
Therefore, it cannot be assumed that the identification
would lie closer to the brighter lobe.
Furthermore, Fig.~\ref{distHist}(e) confirms that while the
optical counterpart is most likely to lie close to the midpoint of the
lobes, the tail of the distribution is not biased towards either lobe.
We used this information extensively in Section~\ref{notes} in 
order to resolve uncertain identifications.

There are two clear outliers in the NVSS centroid histogram (Fig.~\ref{distHist}(a))
with distances to the centroid more than $2.7\sigma$ from the mean. The first,
NVSS~J142312$-$382724 was discussed in Section~\ref{notes}. The second is
NVSS~J024811$-$335106 which is a 33.2-arcsec double. The lobes
are uneven in strength and both show backflow (see Fig.~\ref{overlays}); 
the NVSS centroid lies on the radio axis but much closer to the
brighter lobe than the midpoint. The 
identification is only 2.0\,arcsec
from the midpoint of the lobes and is 0.6\,arcsec from the radio lobe axis.
We are very confident of the identification.

In Fig.~\ref{distHist}(f), 79 per cent of the double sources have identifications
within 1\,arcsec of the axis joining the lobes. All of the sources in 
Fig.~\ref{distHist}(d) that
are more than 0.7\,arcsec from the radio axis have lobe separations 
$>14$\,arcsec, and the median distance from the radio axis for lobe
separations of $>14$\,arcsec is 0.81\,arcsec 
compared to 0.45\,arcsec for those with lobe separations $<14$\,arcsec. 
This indicates that the larger sources are more likely to have complex 
morphologies which can affect the 
measurements of lobe positions.

\subsection{Alignment effect}
\label{pa}

The nature of the link between the presence of a supermassive black hole
and the evolution of its massive elliptical galaxy host remains uncertain.
However, a relationship among the AGN, radio jets and the host galaxy has
been inferred from the `alignment effect' in which the position angle of
the major axis of an elliptical host galaxy has been found to align with
the radio jet axis \citep{McC87,Cha87}.
Alignments are also seen in near-infrared USS-selected HzRG
samples \citep{deB02}, and in high resolution HST images of
HzRGs in the optical \citep[e.g.][]{Bes96} 
and near-infrared \citep[e.g.][]{Pen01,Zir03}.
In contrast \citet{eal97} 
found no significant alignments
in the B2/6C sample, but their sample was small and their near-infrared
images were not as deep as those above.

At low redshift, near-infrared wavebands sample the evolved stellar
population which is unlikely to be affected by the radio jets. At high
redshift ($z>2.5$), however, the observed near-infrared is dominated by
young stars that may result from the radio jet impacting the dense
interstellar and intergalactic medium surrounding the active nucleus. In
the early universe, gas densities in forming galaxies and protoclusters
were much higher than in the nearby universe and shocks associated with
the radio jets are more likely to have triggered star formation along
their path \citep{Bic00,kla04,Cro06}. \citet{vanB98} found a strong evolution with redshift
in the near-infrared alignments of radio host galaxies, with smoother,
less elongated host galaxy morphologies occurring at $z<3$.  In contrast,
\citet{Pen01} found no difference between radio-infrared
alignments in $z<3$ and $z>3$ galaxies.  The absence of alignments in the
\citet{eal97} B2/6C sample, and their clear presence in the 3C
sample \citep{dun93}, led \citet{eal97} to argue that
alignments occur only in the most luminous radio sources.  While 
\citet{deB02} disagree with that interpretation, we will consider in \citetalias{Bry08} 
the luminosity dependence of alignments for those sources with
measured redshifts.

We now present an analysis of radio-infrared alignments for the
MRCR--SUMSS sample.  While such an analysis would ideally be done at the
highest possible spatial resolution, the number of HzRGs that can be
studied with HST is small. The statistical advantage of investigating
alignments in the MRCR--SUMSS sample is the large sample size. Position
angles of the host galaxy were measured on the smoothed $K$-band images by
fitting ellipsoidal profiles using the {\sc iraf/stsdas} task {\sc
ellipse} and taking the major axis as the position angle of the host.  
The position angle, ellipticity and centre of the ellipse were allowed to
vary, and ellipses were fitted as a function of radius. We typically found
that a central ellipse could be fitted with a defined position angle, but
as the radius increased there were one or two sudden steps in the position
angle. In these cases, we took the outermost radius at which an ellipse
could be fitted with a stable ellipticity and position angle. For some
images the position angle was constantly changing with radius and no
stable ellipse could be fitted.
The galaxies that could not be fitted
included those with low $K$-band signal-to-noise ratio, and those with
clumpy, circular or disturbed morphologies.

\begin{figure}
\centerline{\psfig{file=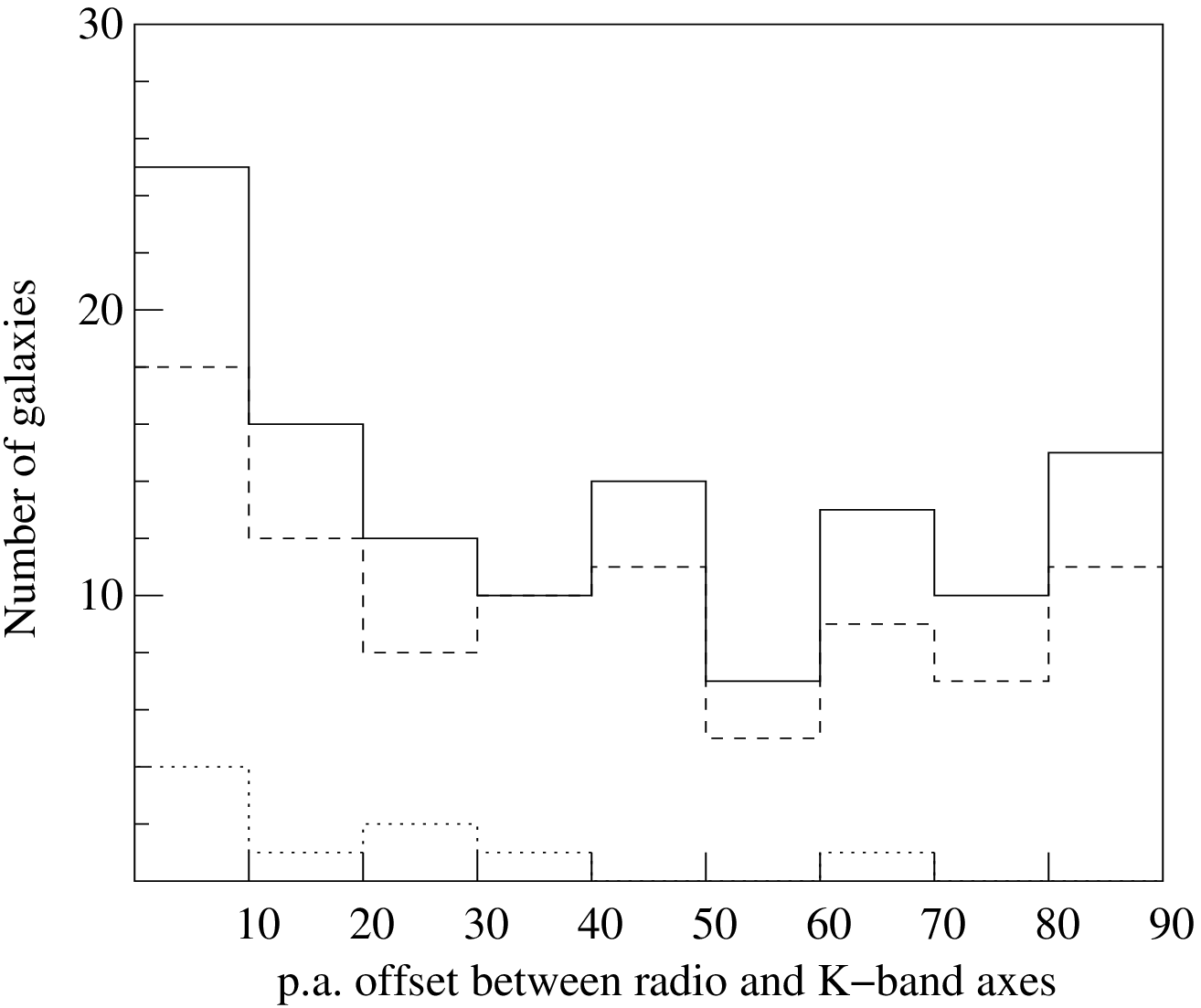,width=6.5cm}}
 \caption{Distribution of differences between the radio axis position
angle and the major axis of the $K$-band fitted ellipsoid for 122 galaxies
from the MRCR--SUMSS sample (solid line) and for those with largest
angular size greater than five arcsec (dashed line). The dotted line shows
the difference between the radio position angle and that of the extended
$K$-band emission for the nine sources which have this feature. }
 \label{pa_hist}
\end{figure}

Of the 152 sources with PANIC or IRIS2 $K$-band imaging, we fitted position angles to 131
$K$-band images, of which nine did not have a well defined radio axis.  For the
remaining 122 sources we measured the offset between $K$-band and radio
position angles. 
Fig.~\ref{pa_hist} shows the
distribution of differences between the position angle of the radio axis
and the major axis of the fitted $K$-band ellipsoid. While some samples in
the literature show alignments over angles from 0--30
degrees \citep[e.g.][]{deB02} we find a more uniform distribution
of offset angles, with a small excess of alignments within 10 degrees.
This excess suggests that the process causing the alignment of the
outermost isophotes of the host galaxy is tightly collimated along the
radio jet axis, as might be expected for jet-induced star formation.  
Alternatively, the small alignment angles could be the result of radio
jets having broken through gas clouds along their path, leaving an
increased surface area of cooled gas for scattering light from the active
nucleus \citep{Bre97}.

The sources in our sample with LAS\ $>$\ 5\,arcsec have the same
percentage (21 per cent) aligned within ten degrees as in the complete
sample. This is in contrast to other papers \citep[e.g.][]{deB02}
where the larger sources have been found to be more strongly aligned. To
test whether the ellipsoidal fitting may have missed faint emission with
low signal-to-noise, we rotated all the smoothed $K$-band images to align
with the radio position angles. The images were then co-added after
careful scaling by the background level and the galaxy peak brightness.
The result showed no ellipticity, confirming that alignments are not
present in the bulk of our sample. As most of the sources in our sample
are at $z<3$, this result is consistent with the ground-based sample of
\citet{vanB98} but in stark contrast to \citet{Pen01} who found a 
predominance of aligned features in their near-infrared
HST images.  We attribute this difference to the different surface
brightness sensitivities and resolutions of HST and ground-based images.

Nine galaxies in our sample (listed in Table~\ref{pa_galaxies}) show
$K$-band emission extending beyond the host galaxy ellipsoid. In these
cases the overall position angle was estimated by eye, and the
distribution of offsets from the radio axis is shown in
Fig.~\ref{pa_hist}. Four of the nine have position angles within
10\,degrees of the radio position angle, and seven are within 30\,degrees.  
The origin of these alignments could again be the result of jet-induced
star formation or scattered nuclear light along the path of the radio
jets. The misaligned cases may be due to changes in jet direction, or the 
chance projection of other objects close to the line of sight to the host 
galaxy.

\begin{table*}
\caption{Position angle offsets between the radio axis and, firstly, the extended $K$-band emission 
($\delta$PA$_{\rm ext}$) and, secondly, the major axis of the host ellipsoid ($\delta$PA$_{\rm ellipse}$) for the galaxies
that have extended $K$-band structure extending beyond the host galaxy ellipsoid.
}
\label{pa_galaxies}
\begin{center}
\begin{tabular}{lccc}
\hline
\hline
Object & $\delta$PA$_{\rm ext}$ & $\delta$PA$_{\rm ellipse}$\\
  & (degrees)  &    (degrees)   \\
\hline
NVSS~J004000$-$303333 & 19 & 42 \\
NVSS~J004030$-$304753 & 22 & 34 \\
NVSS~J021208$-$343111 & 34 & 0.8 \\
NVSS~J140010$-$374240 & 62 & 59 \\
NVSS~J144932$-$385657 & 8 & 31 \\
NVSS~J151020$-$352803 & 5 & 15 \\
NVSS~J215047$-$343616 & 9 & 0.5 \\
NVSS~J221650$-$341008 & 23 & 30 \\
NVSS~J234235$-$384526 & 5 & 15 \\
\hline
\end{tabular}
\end{center}
\end{table*}

NVSS~J111921$-$363139 (see Fig.~\ref{highRes}) has extended $K$-band
emission which can not be fitted by an ellipsoid, with broad patches of
emission lined up along the radio axis. This morphology has been
previously associated with either ionisation cones or the presence of dust
in the host galaxy.  At the redshift of the host galaxy ($z=2.769$) there
are no prominent emission lines in the $Ks$-band filter, so we may be
seeing either an inclined dust lane, through which the nucleus can be seen
\citep[similar to ESO 248-G10,][]{Bry00}, 
or widely distributed
dust and star formation along the radio jet axis.

\subsection{Aperture corrections}

\begin{figure}
\centerline{\psfig{file=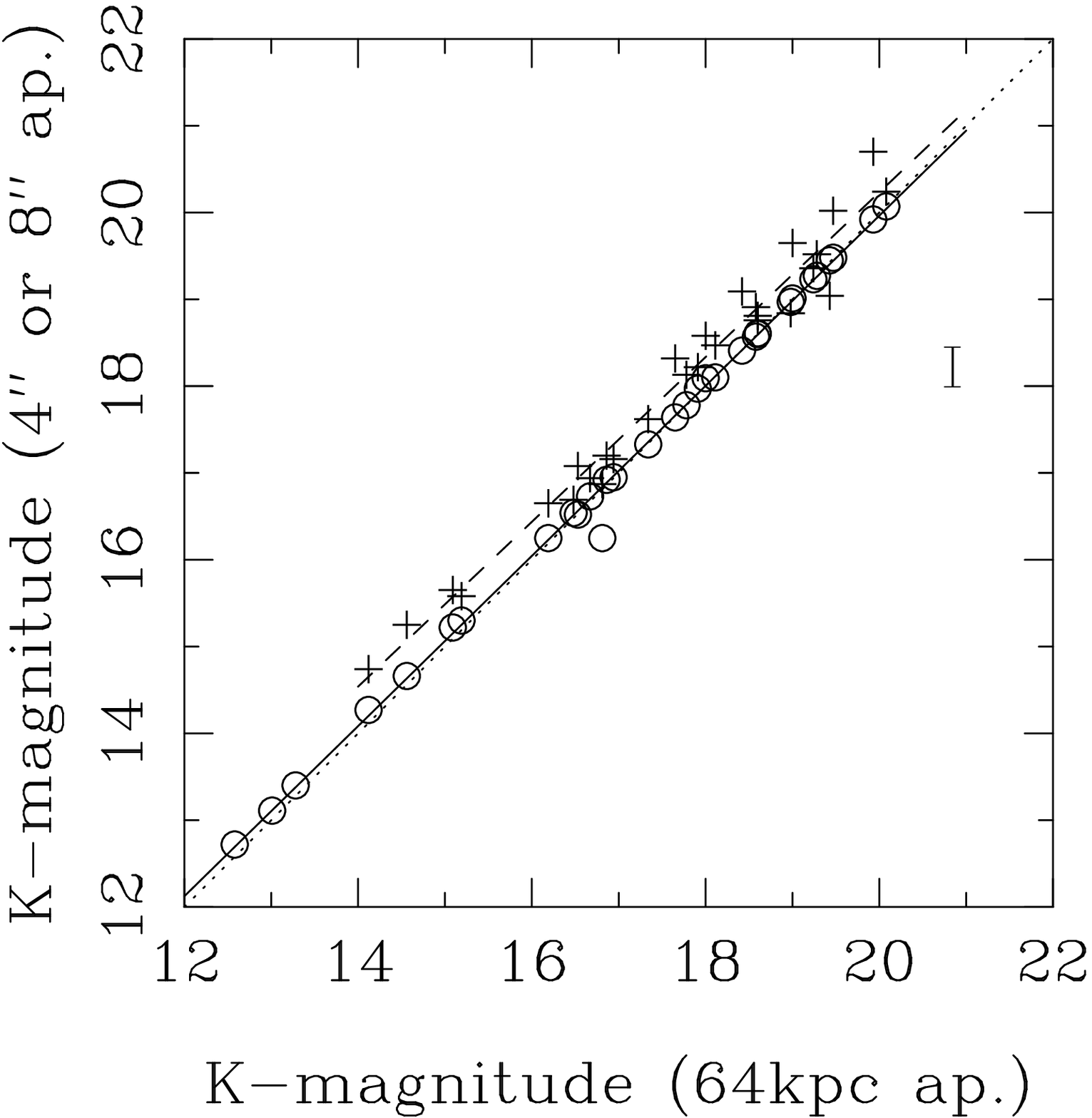,width=6.5cm}}
\caption{$K$-band magnitudes corrected to 64-kpc apertures versus 8-arcsec (circles) and 4-arcsec (+)
apertures from the SUMSS--NVSS data set \citep{deB04,deB06}.
The solid line
is the linear fit to the 8-arcsec points, the dashed line is the fit to the
4-arcsec-aperture points, while the dotted line is the line of equality. An error
bar represents the average error in the 8-arcsec-aperture $K$-band magnitudes. It
is clear that the difference between the 8-arcsec- and 64-kpc-aperture 
$K-$magnitudes is small compared to errors. The 4-arcsec-aperture values follow a
very similar gradient, with a fit equation $K {\rm (64\,kpc)} = 1.026*{\rm log}[K(4'')] - 0.824$.
}
\label{KK}
\end{figure}

In some earlier papers \citep[e.g.][]{eal97,Wil03,deB04}
the $K$-band magnitudes have been converted to an
equivalent 64-kpc aperture value. This value was chosen by \citet{eal97} 
because it corresponds to an 8-arcsec aperture at $z=1$.
Fig.~\ref{KK}, based on data from \citet{deB06}, shows that the
actual correction to the $K$-band magnitudes is quite small and applies
over the full magnitude and redshift range. The basis of the empirical
correction adopted by \citet{eal97} is that the integrated emission
of galaxies at $z>0.6$ within an aperture of radius $r$ is proportional to
$r^{0.35}$. While a profile of this form may be an acceptable fit to
nearby, bright galaxies, the galaxies in our sample are faint and distant,
and the $K-$band images are not deep enough to justify fitting such a
profile. 

At high redshift ($z>2$) it is expected that many galaxies are still in
the process of merging and have not yet relaxed to an elliptical profile
\citep[e.g.][]{vanB98,Pen98}.  An $r^{0.35}$
profile is therefore unlikely to apply over an epoch range as large as
$z=0.6$--$4$. This is supported by \citet{vanB98} who found
that the surface brightness profiles of $z<3$ HzRGs are steeper than those
at $z>3$. The assumption of a `universal' radial profile then introduces
an added error, estimated to be as large as 20 per cent.  A contribution
to this error comes from the chosen starting aperture which could, in
principle, be based on the profile of the galaxy.  The practical issue,
however, is that profiles can not be determined accurately for the faint
and distant galaxies.  As a result, some authors have adopted an 8-arcsec
aperture from which to convert to a 64-kpc aperture based on the
simplicity of their equivalence at $z=1$. In practice an 8-arcsec aperture
is often too large to avoid including additional flux from nearby
galaxies. \citet[][fig.  1]{deB02} 
showed that the majority of
HzRGs in their sample have most of their flux contained within a 2-arcsec
aperture.

\begin{figure*}
\begin{minipage}[]{0.48\textwidth}
\centerline{\psfig{file=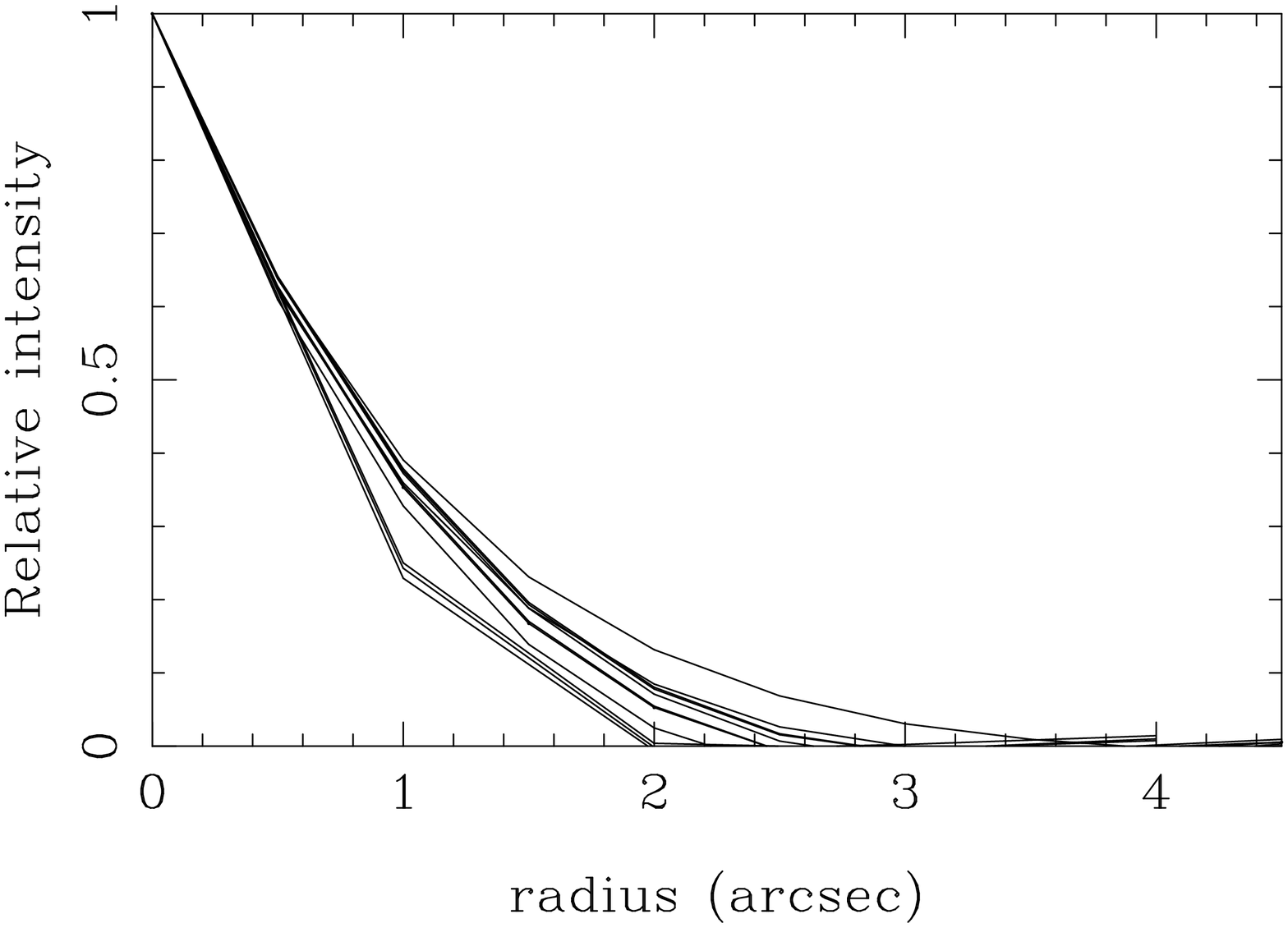,width=8.0cm}}
\end{minipage}%
\hspace*{4mm}
\begin{minipage}[]{0.48\textwidth}
\centerline{\psfig{file=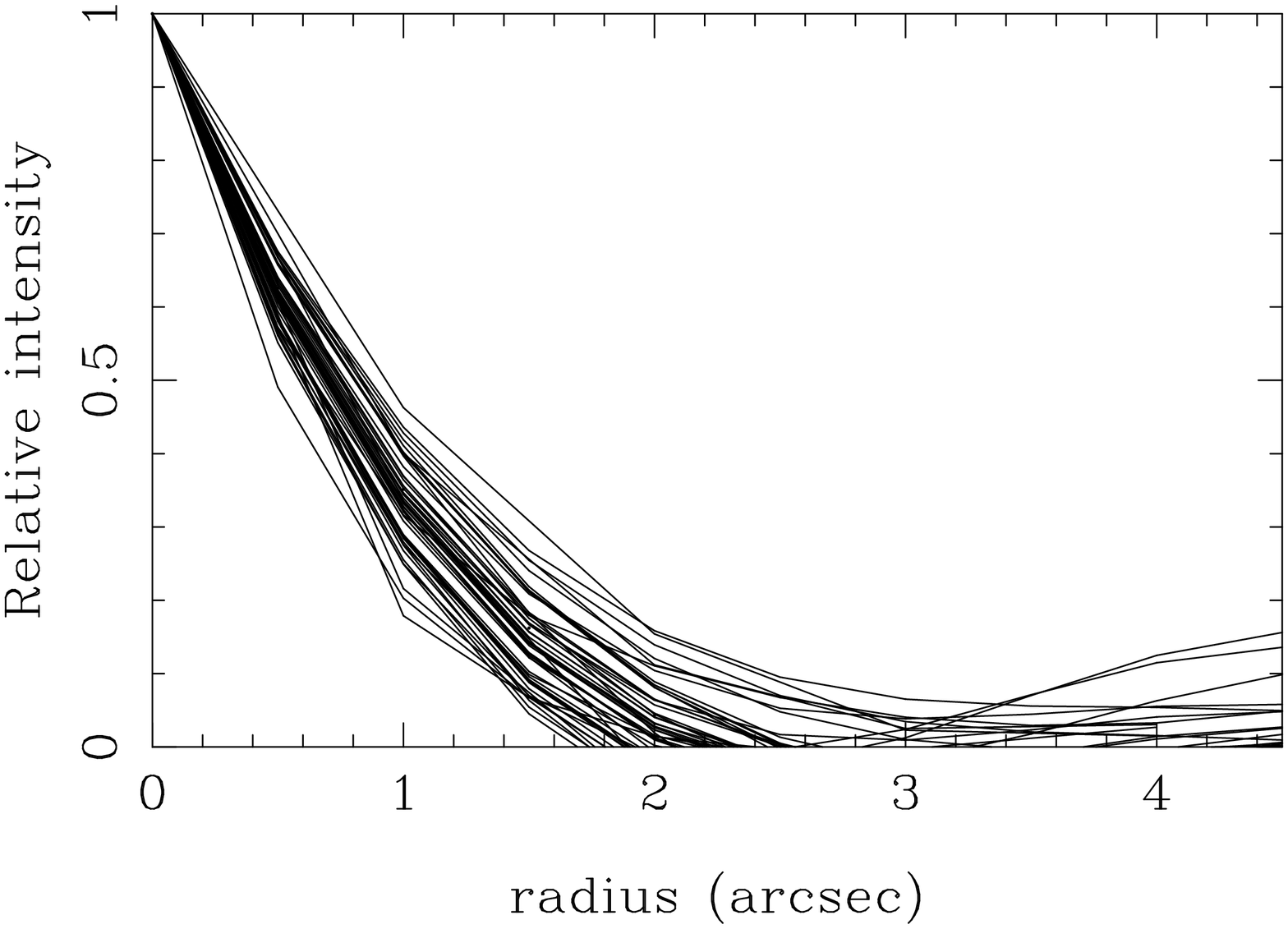,width=8.0cm}}
\end{minipage}
\caption{Relative intensity at each radial step is plotted against radial steps
of 0.5\,arcsec for all of the PANIC and IRIS2 images with $K<17$ (left) and
$K>19$ (right). In almost every case the emission from the galaxy is
contained within a 4-arcsec-diameter aperture, and in some cases, emission from nearby 
objects is beginning to contaminate an 8-arcsec-diameter aperture.}
\label{profiles}
\end{figure*}

In Fig.~\ref{profiles} we plot the radial profiles of our $K$-band
identifications with $K>19$ and $K<17$ representing nominal high- and
low-redshift groups respectively. Almost all the galaxies have their total
emission enclosed by a 4-arcsec aperture, with no difference between the
brighter and fainter sources. Infrared imaging is relatively insensitive
to low-surface-brightness emission, and for most of our objects, the flux
has dropped to the background sky level within a 4-arcsec aperture.  On
the other hand, there are frequent cases where the intensity increases
again beyond a radius of 2 arcsec, indicating contaminating emission from
a nearby source. For the sources in our sample, therefore, we do not
believe there is any significant missing flux in the 4-arcsec apertures.  
Extending the aperture to 8-arcsec adds sky noise, which increases the
error.

Over the large time interval from $z\sim4$ to the present, galaxies
undergo significant evolution. We have no evidence that one physical size
scale is a reasonable representation of HzRGs which may range over a
factor of ten in mass \citep{roc04} and which we observe
at different evolutionary stages in their formation. For these reasons, we
have chosen not to use 64-kpc-aperture magnitudes, and instead to use
4-arcsec-aperture magnitudes in this paper.

\subsection{$K$ and flux density distributions}

\begin{figure*}
\psfig{file=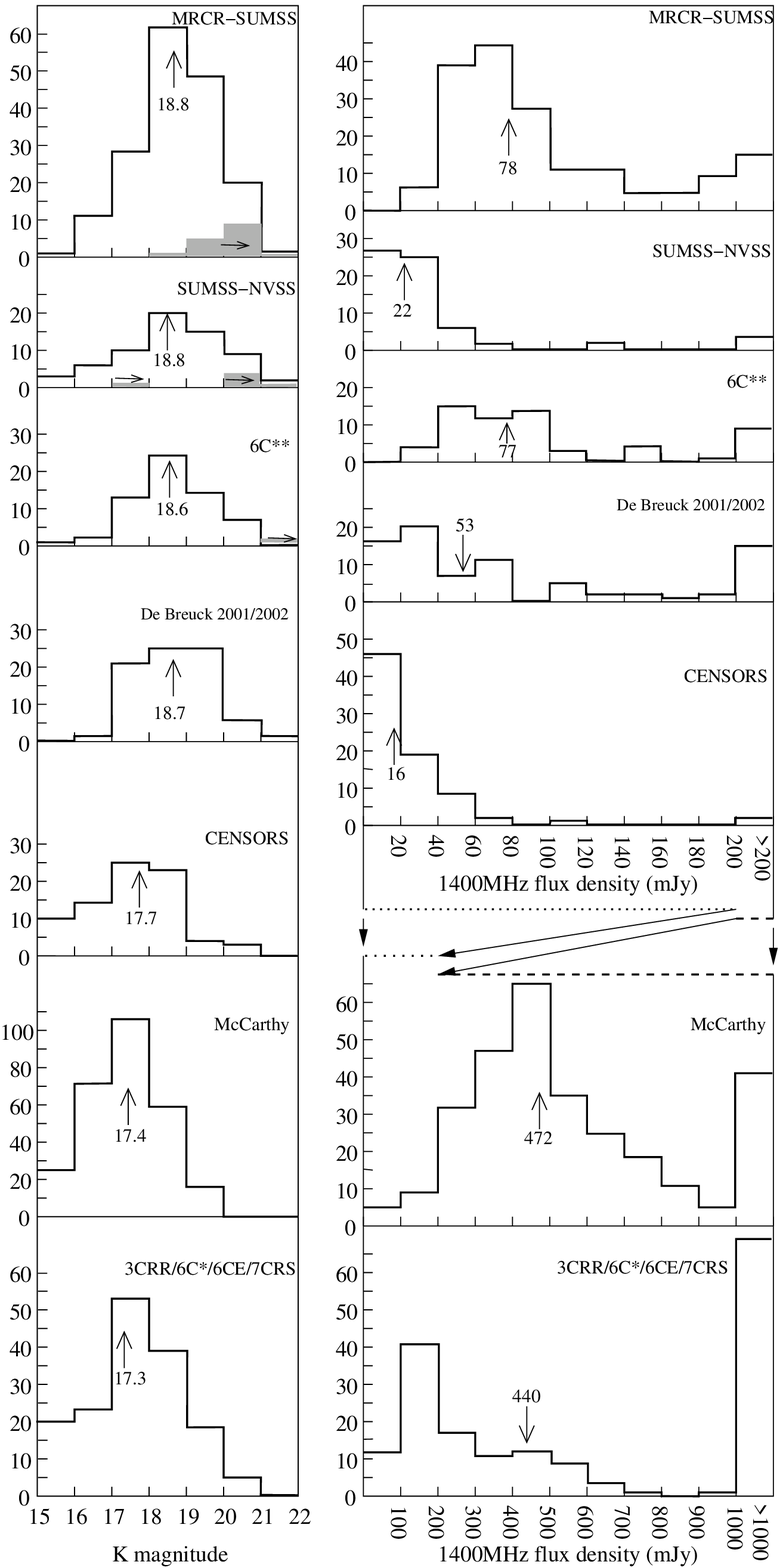, width=8.5cm}
\caption{Distributions of NVSS-1400-MHz-flux densities and $K$-band magnitudes for the
samples from MRCR--SUMSS, SUMSS--NVSS \citep{deB04,deB06}, \citet{deB01,deB02},
6C** \citep{cru06}, CENSORS 
\citep{bro06,bro08},
\citetalias{McC96}
and a combined sample from the 3CRR, 6C$^{*}$, 6CE and 7CRS catalogues \citep[compiled by][]{Wil03}.
The 4-arcsec-aperture values for the MRCR--SUMSS, SUMSS--NVSS and \citet{deB01,deB02}
samples are shown, while the plots for the 6C** and CENSORS samples show
4-arcsec-aperture magnitudes calculated from the published 3- and 5-arcsec values
by averaging the equivalent 3- and 5-arcsec fluxes and converting to a 4-arcsec magnitude. 
The McCarthy magnitudes are in a
3-arcsec aperture and based on our dataset, we expect the average offset between the
3-arcsec and 4-arcsec aperture magnitudes to be $<0.2$. 
The combined 3CRR/6C$^{*}$/6CE/7CRS sample from \citet{Wil03} has equivalent 
64-kpc aperture $K$-band 
magnitudes, which are expected to be an average of 0.4 magnitudes fainter than the
equivalent 4-arcsec-aperture magnitudes for those objects (see Fig.~\ref{KK}).
Vertical arrows mark the median values in each
plot, calculated from the confirmed values and excluding lower limit
$K$ magnitudes shown by the shaded regions. 
The non-USS samples contain sources brighter than 
$K=15$, and they are included in the median calculation. The CENSORS sample has one source
with 5 $K$-band components and hence there are 4 fewer sources in the flux density histogram
than in the $K$-band histogram. The 3CRR/6C$^{*}$/6CE/7CRS sample has 26 sources that
do not have published 1400-MHz flux densities and are 
included in the
$K$-band histogram, but not the flux density histogram.
}
\label{K_Shist}
\end{figure*}

Essentially all USS-selected HzRG samples have selection biases 
resulting from the different spectral index
cut-offs and selection frequencies, which results in samples with vastly
different flux density distributions. A k-correction
was thought to underpin the effectiveness of USS-selection as a steepening radio SED would 
be shifted with redshift to lower frequencies and therefore the highest redshift objects 
would be steepest at a given observed frequency \citep{Gop88,Car99}. Therefore
USS samples selected at lower frequencies should net a higher redshift
distribution.
This theory was challenged by \citet{kla06} who found that the SEDs of sources in the 
843-MHz-selected SUMSS--NVSS
sample are predominantly straight. 
The MRCR--SUMSS sample was then
selected at the lower frequency of 408\,MHz, and the resulting $K$-band magnitude and 1400\,MHz
flux density distributions are compared in Fig.~\ref{K_Shist}.
The other samples shown are USS-selected samples with 4-arcsec-aperture
magnitudes including the SUMSS--NVSS, 6C** \citep{cru06} and \citet{deB01,deB02} 
samples, along with the three non-USS-selected samples from 
\citetalias{McC96}, 
CENSORS \citep{bro06,bro08} and a combined sample from the 3CRR, 6C$^{*}$, 6CE and 7CRS catalogues \citep[compiled by][]{Wil03}.
The McCarthy sample is a set of 277 $K$-band observations of radio
galaxies, of which 175 have redshifts. While it remains unpublished,
it is the largest non-USS $K$-band sample and therefore provides an 
interesting comparison
with the other samples discussed here. We obtained 1400\,MHz flux density values
for the McCarthy sample by cross-matching the positions with the NVSS catalogue 
within a 60-arcsec search radius. For the 38 of the 293 sources which had
two matches within 60\,arcsec, we selected the closest match.
SUMSS--NVSS was selected to have $\alpha_{843}^{1400}<-1.3$, and
the 6C** sample is from the
151-MHz 6C survey with $\alpha_{151}^{1400}<-1.0$.
The \citet{deB01,deB02} 
sample was selected at one of the
frequencies 325, 365, or 408\,MHz as detailed in the papers,
and has $\alpha_{\sim350}^{1400} \lesssim -1.3$. 
The combined sample compiled by \citet{Wil03} includes 202 narrow line galaxies
from the 3CRR, 6C$^{*}$, 6CE and 7CRS catalogues, selected at 151 and 178\,MHz. 
The 1400\,MHz flux densities were 
obtained from the NASA/IPAC Extragalactic Database (NED) by selecting the most 
recently published\footnote{Flux densities came from \citet{Con98}, \citet{Lai80}, \citet{Kel69}, \citet{Whi92}, \citet{Owe97}, \citet{Cro05} and \citet{Bec95}} total integrated flux density. In several cases only peak
flux densities were available and 26 targets have no published observations at 
1400\,MHz. While the 3CRR/6C$^{*}$/6CE/7CRS combined sample is less complete in 1400\,MHz 
flux density than the other samples shown in Fig.~\ref{K_Shist}, it has a 
more complete redshift follow-up (see Paper\,III).

The four USS samples have very similar $K$-magnitude distributions, which may
suggest that when the redshift followup is complete, the redshift distributions
may also be similar.
A Kolmogorov-Smirnov
test comparing the $K$ distributions of the other samples with the MRCR--SUMSS
sample, suggests that the three other USS-selected samples (SUMSS--NVSS, 6C** and
De Breuck 2001, 2002) are likely to be drawn from the same distribution as the MRCR--SUMSS data, with
$p$ values of 0.57, 0.75 and 0.59. However, the three non-USS-selected samples
(CENSORS, 3CRR/6C$^{*}$/6CE/7CRS and McCarthy) are not from the same distribution, with $p<0.0005$. The MRCR--SUMSS
sample is much larger than the other USS-selected samples in terms of the
number of $K$ measurements, and therefore will give a more complete picture
of USS radio galaxy redshift distributions once the spectroscopic followup
is complete.

The most striking result is that even though the USS-selected samples each have
completely different selection criteria and vary by a factor of 3.5 in median flux
density, the resultant $K$-band distributions are the same, and are fainter
than for the non-USS-selected samples. The non-USS-selected samples have vastly
different flux density distributions, but have similar 
$K$-band medians that are brighter than the USS-selected samples. Therefore the effectiveness of
USS-selection in identifying fainter and potentially higher-redshift galaxies 
in the samples shown, 
is not dependent on the 
selection frequency or the resultant 
flux density distributions of these samples.
Based on the $K-z$ relation, if this
result also applies to the redshift distribution, then it will provide strong evidence
against the k-correction theory which has been thought to account for the
effectiveness of USS-selection.

\section{CONCLUSIONS}

$K$-band imaging has been presented for a large Southern
hemisphere USS-selected HzRG survey, the MRCR--SUMSS sample (408--843\,MHz). 
We have the following findings:

(1) Based on
163 new $K$ images, 152 radio sources (93 per cent) have $K$-band identifications
with the faintest at $K=21.4$. In combination with the 12 previously-published
observations, 175 targets have followup observations and 94 per cent of these have
detections. This is consistent
with the 96 per cent detection rate of all $z>2$ HzRGs to $K=22$ in the literature \citep{Mil:08}.

(2) We have investigated the location of the identifications in relation to the
radio structures for double sources and found that the midpoint of the lobes is the most
likely place to find the identification, with 66 per cent of counterparts 
located within 2\,arcsec of the midpoint.
The radio centroid is a less likely position to find the
identification.
We have used this analysis as a guide when selecting identifications in
cases where there were ambiguities. 

(3) We find that the identification is no more
likely to be closer to the brighter lobe than the fainter one, even
though selection of the identification by eye may favour an
identification closer to the brighter lobe. While 79 per cent of identifications are within 1\,arcsec of the axis joining
the lobes, the off-axis distance increases for sources with larger
lobe separations.

(4) The size of the MRCR-SUMSS sample is larger than most previous samples that have
investigated the alignment effect between radio and near-infrared axes. We have found that
21 per cent of the galaxies that could be fitted with an ellipsoid, have a major
axis that aligns within 10\, degrees of the radio axis. There is no evidence for
alignment at angles larger than 10\,degrees, lending support to jet-induced-star-formation
as the origin of the alignment. We identify non-elliptical, extended aligned structures in
nine galaxies that cannot all be explained by jet-induced-star-formation. 

(5) By comparing the MRCR--SUMSS sample $K$-band and 1400\,MHz flux density distributions
with those of other USS-selected and non-USS-selected samples, we find that while the flux
density distributions vary significantly between samples due to different selection
criteria, the median $K$-band magnitudes are similar
for all the USS-selected samples. If the success of USS-selection is due
to a k-correction, we would expect the samples selected at lower frequency to have
a fainter magnitude distribution. This needs to be tested further with comparative
redshift distributions.

\citetalias{Bry08} will present spectroscopy for this sample and interpret the radio and $K$-band results in
terms of the redshifts found.

\label{lastpage}

\subsection*{Acknowledgements}
We acknowledge financial support from the {\it Access to Major Research Facilities
Programme} which is a component of the {\it International Science Linkages Programme} established under the Australian Government's innovation
statement, {\it Backing Australia's Ability}.

Australian access to the Magellan Telescopes was supported through the 
Major National Research Facilities program of the Australian Federal Government.

JWB acknowledges the receipt of both an Australian Postgraduate Award and a 
Denison Merit Award. RWH, HMJ and JJB acknowledge support from the Australian 
Research Council and the University of Sydney Bridging Support Grants Scheme. 
BMG acknowledges the support of a Federation Fellowship from the Australian 
Research Council through grant FF0561298.

We thank Elaine Sadler for many useful discussions, Pat McCarthy for providing his 
unpublished data sample, David Crawford 
for providing the MRCR, Gemma Anderson for doing some observations for us,
and the team at the Magellan Telescopes for their exceptional efficiency and
good humour. We also thank the referees for their valuable comments which improved the paper. 

The Australia Telescope Compact Array is part of the 
Australia Telescope which is funded by the Commonwealth of Australia for 
operation as a National Facility managed by CSIRO. SuperCOSMOS Sky Survey 
material is based on photographic data originating from the UK, Palomar and 
ESO Schmidt telescopes and is provided by the Wide-Field Astronomy Unit, 
Institute for Astronomy, University of Edinburgh. This research has made use 
of the NASA/IPAC Extragalactic Database (NED) which is operated by the Jet 
Propulsion Laboratory, California Institute of Technology, under contract 
with the National Aeronautics and Space Administration.
IRAF is distributed by the National Optical Astronomy Observatories, 
which are operated by the Association of Universities for Research 
in Astronomy, Inc., under cooperative agreement with the National 
Science Foundation.
This publication makes use of data products from the Two Micron 
All Sky Survey, which is a joint project of the University of 
Massachusetts and the Infrared Processing and Analysis Center/California Institute of Technology, funded by the National Aeronautics and
Space Administration and the National Science Foundation.

\newpage
\newpage

\renewcommand\thefigure{A-1}
\newpage
\section*{APPENDIX}
\begin{figure*}
\begin{center}
\psfig{file=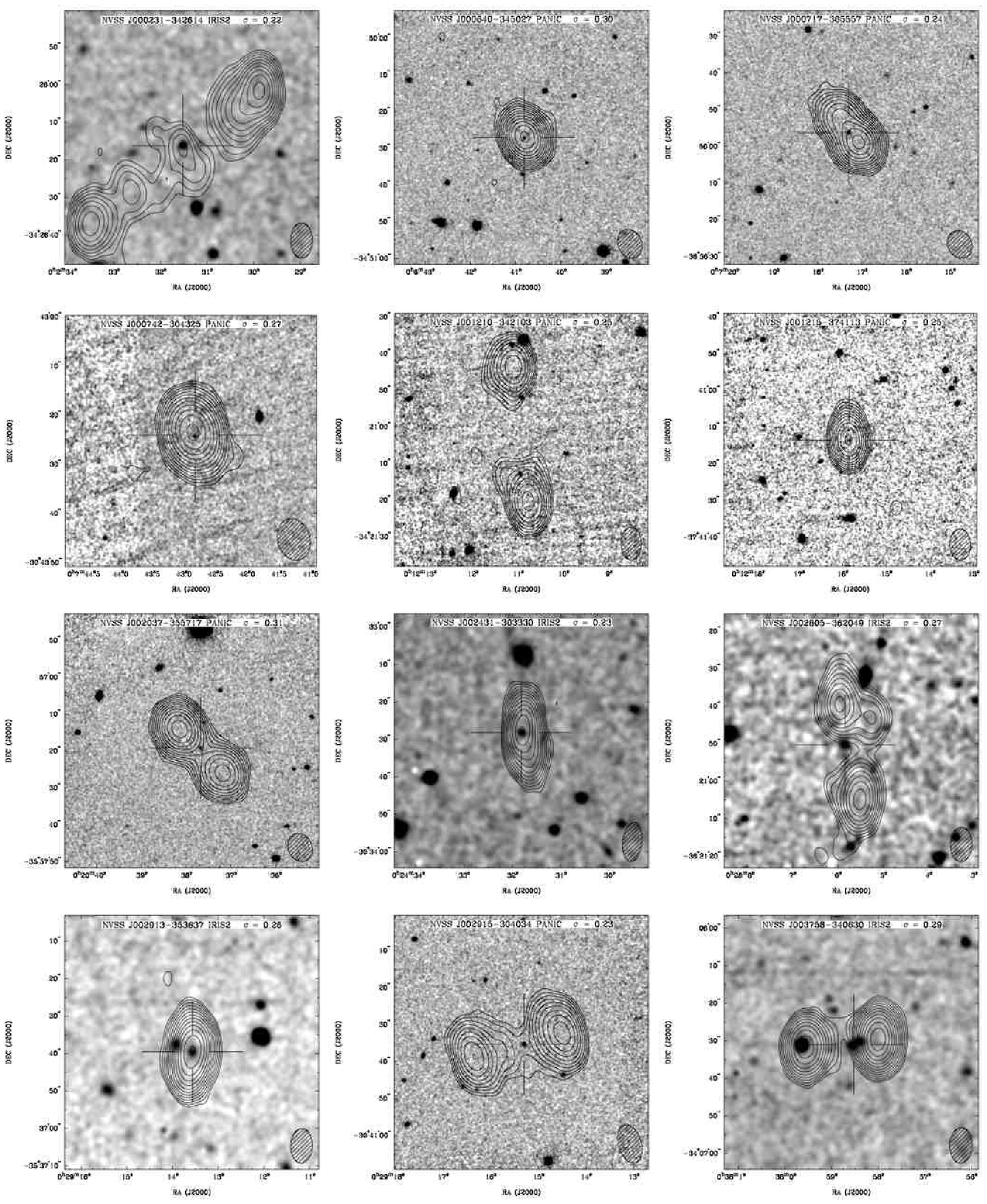,bbllx=50pt,bblly=20pt,bburx=570pt,bbury=800pt,width=17.35cm,clip=}
\vspace{-5cm}
\caption{2368-MHz ATCA contours overlaid on the $K$-band images for
all the sources in our sample that have IRIS2 or PANIC images and
are not shown in Fig.~\ref{overlays}. The full set of Appendix images are 
included in the MNRAS online journal. The $K$-band images show
which instrument they are from, in the header,
and they have been smoothed using a gaussian kernel of 3 pixels FWHM.
All the radio contours are from natural-weighted images.
The lowest contour is 3 sigma, and the contours
are a geometric progression in $\sqrt 2$. The rms noise ($\sigma$) is shown in the header of each
image in mJy\,beam$^{-1}$.
Crosshairs mark the $K$-band counterpart to the radio source.
that were identified in $K$.
The ATCA synthesized beam is shown in the bottom right-hand corner of each panel.
}
\label{app_overlays}
\end{center}
\end{figure*}

\renewcommand\thefigure{A-2}

\begin{figure*}
\begin{center}
\psfig{file=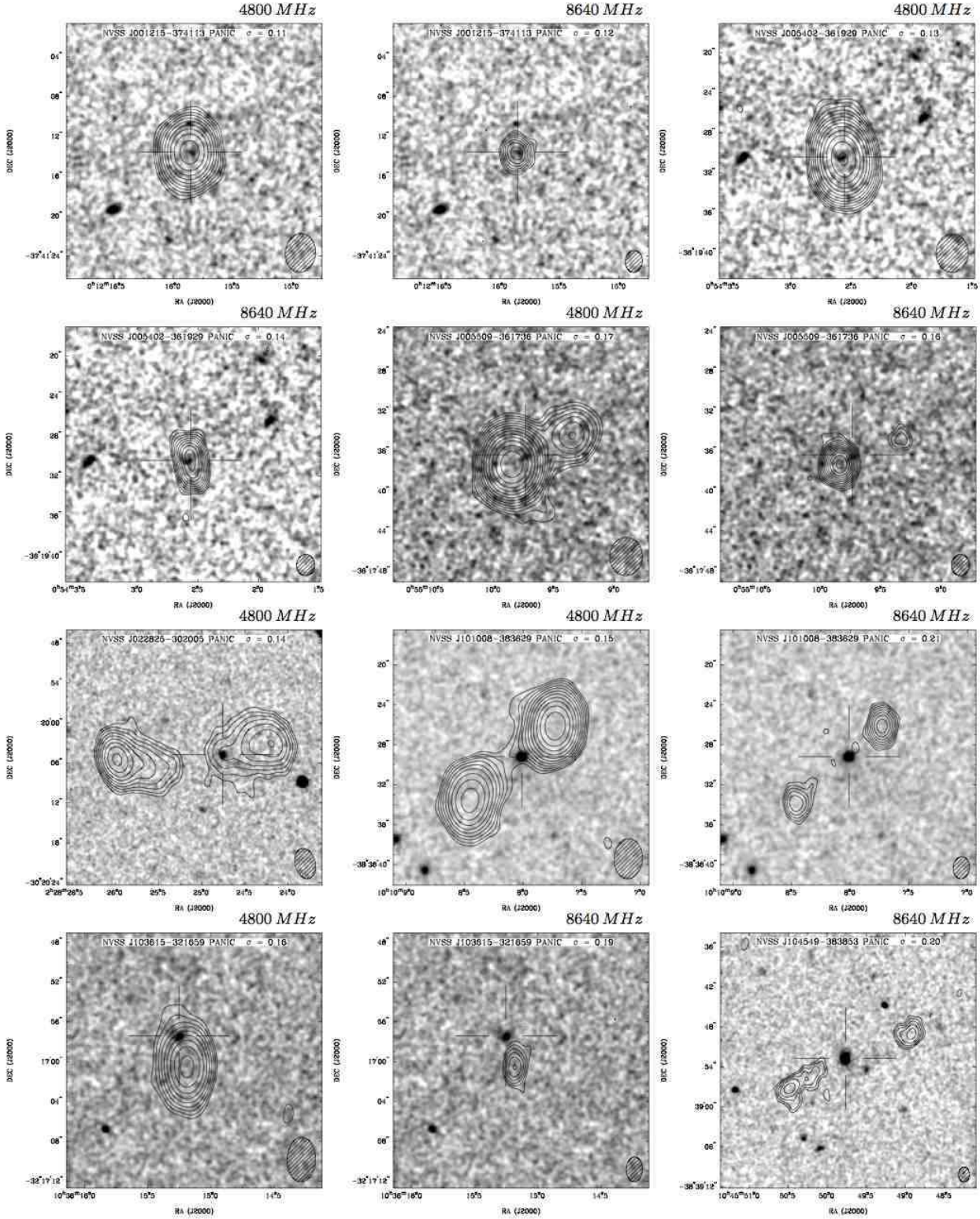,bbllx=50pt,bblly=20pt,bburx=570pt,bbury=800pt,width=17.2cm,clip=}
\vspace{-5cm}
\caption{
$K$-band images for
all the sources in our sample that have ATCA images at 4800- or 8640-MHz including all the images not already shown
in Fig.~\ref{highRes}. The full set of Appendix images are included in the 
MNRAS online journal. The frequency of the
overlaid contours is marked on the 
top right of each image.
The $K$-band images show
which instrument they are from, in the header,
and they have been smoothed using a gaussian kernel of 3 pixels FWHM.
All the radio contours are from natural-weighted images.
The lowest contour is 3 sigma, and the contours
are a geometric progression in $\sqrt 2$. The rms noise ($\sigma$) is shown in the header of each
image in mJy\,beam$^{-1}$.
Crosshairs mark the $K$-band counterpart to the radio source.
that were identified in $K$.
The ATCA synthesized beam is shown in the bottom right-hand corner of each panel.
}
\label{appendixHR}
\end{center}
\end{figure*}

\end{document}